\documentclass[notoc,hyper,letterpaper]{JHEP3}

\usepackage{epsfig}
\usepackage{amsbsy}
\usepackage{varioref}
\usepackage{pifont}

\newcommand{\dfrac}[2]{\frac{\strut \displaystyle{#1}}
                      {\strut \displaystyle{#2}}}

\title{The Structure of Corrections to Electroweak\\
Interactions in Higgsless Models}

\author{R. Sekhar Chivukula and Elizabeth H. Simmons\\
Department of Physics and Astronomy, Michigan State University\\
East Lansing, MI 48824, USA\\
	E-mail: \email{sekhar@msu.edu, esimmons@msu.edu}}
	
\author{Hong-Jian He\\
Department of Physics, University of Texas\\
Austin, TX 78712, USA\\
	E-mail: \email{hjhe@physics.utexas.edu}}

\author{Masafumi Kurachi\\
Department of Physics, Nagoya University\\
Nagoya 464-8602, Japan\\
	E-mail:\email{kurachi@eken.phys.nagoya-u.ac.jp}}

\author{Masaharu Tanabashi\\
Department of Physics, Tohoku University\\
Sendai 980-8578, Japan\\
	E-mail:\email{tanabash@tuhep.phys.tohoku.ac.jp}}

\abstract{
Recently,  ``Higgsless'' models of electroweak symmetry breaking have
been proposed. Based on compactified five-dimensional gauge theories, these models achieve unitarity of electroweak boson self-interactions 
through the exchange of a tower of massive 
vector bosons rather than the exchange of a scalar Higgs boson. In this paper, using
deconstruction, we analyze the form of the corrections to the electroweak interactions in a large class of these models, allowing for arbitrary 5-D geometry, position-dependent 
gauge coupling, and brane kinetic energy terms. We show that many models considered in the literature, including those most likely to be
phenomenologically viable, are in this class. By analyzing the asymptotic behavior of
the correlation function of gauge currents at high momentum, we extract the exact form of the relevant correlation functions at tree-level and compute the
corrections to precision electroweak observables in terms of the spectrum
of heavy vector bosons. We determine when nonoblique corrections due to the interactions of fermions with the heavy vector bosons become important, and
specify the form such interactions can take. In particular we find that in 
this class of models, so long
as the theory remains unitary, $S-4 \cos^2\theta_W T
> {\cal O}(1)$, where $S$ and $T$ are the usual oblique parameters.
We concur
with the result of Cacciapaglia {\it et.al.} that small or negative $S$ is possible -- though
only at the expense of substantial negative $T$ at tree-level. 
Although we stress our results as they apply to continuum
5-D models, they apply also to models of extended  electroweak gauge symmetries 
motivated by models of hidden local symmetry. 
}

\preprint{
{MSUHEP-040607} \\
{DPNU-04-12} \\
{TU-722}
}

\begin{document}


\section{Introduction}

The origin of electroweak symmetry breaking remains one of the most significant
questions in elementary particle physics. Recently, ``Higgsless'' models of electroweak symmetry breaking have been proposed \cite{Csaki:2003dt}. 
Based on five-dimensional gauge theories compactified on an interval, these models 
achieve unitarity of electroweak boson self-interactions 
through the exchange of a tower of massive vector bosons 
\cite{SekharChivukula:2001hz,Chivukula:2002ej,Chivukula:2003kq}, 
rather than the exchange of a scalar Higgs boson \cite{Higgs:1964ia}.
Motivated by gauge/gravity duality \cite{Maldacena:1998re,Gubser:1998bc,Witten:1998qj,Aharony:1999ti}, models of this kind may be viewed as ``dual'' to more conventional models of dynamical symmetry breaking 
\cite{Weinberg:1979bn,Susskind:1979ms} such
as ``walking techicolor'' \cite{Holdom:1981rm,Holdom:1985sk,Yamawaki:1986zg,Appelquist:1986an,Appelquist:1987tr,Appelquist:1987fc}, and they represent an exciting new avenue of investigation.

In this paper, using deconstruction  \cite{Arkani-Hamed:2001ca,Hill:2000mu}, we calculate the form of the corrections to the electroweak interactions in a large class of these models.\footnote{To be
precise, in this analysis we are concerned with the corrections to electroweak processes
at tree-level in the additional interactions. These corrections exist independent of
any particular ``high-energy'' completion of the deconstructed theory. 
Recently, Perelstein \protect\cite{Perelstein:2004sc} has concluded that
the effects of a QCD-like high-energy completion -- arising from higher-order
contributions analogous to $L_{10}$ in
the chiral Lagrangian -- are potentially large as well.} 
Our analysis applies to any Higgsless model which can be deconstructed to a  chain of $SU(2)$ gauge groups adjacent to a chain of $U(1)$ gauge groups, 
with the fermions coupled to the $SU(2)$ group at the end of the chain and to the
 $U(1)$ group at the interface between the $SU(2)$ and $U(1)$ chains.
We show that many models considered in the literature, including those most likely to be
phenomenologically viable \cite{Cacciapaglia:2004jz}, are in this class. By analyzing the asymptotic behavior of the correlation function of gauge currents at high momentum
in a deconstructed model of this sort, generalizing an argument for dual
models of QCD \cite{Son:2003et,Hirn:2004ze,Chivukula:2004kg}, we extract the exact form of the relevant correlation functions at tree-level and compute the corrections to precision electroweak observables. We determine when nonoblique corrections due to the interactions of fermions with the heavy vector bosons become important, and
specify the form such interactions can take. In particular we find that in this class of models, so long as the theory remains unitary, $S-4 \cos^2\theta_W T
> {\cal O}(1)$, where $S$ and $T$ are the usual oblique parameters
\cite{Peskin:1992sw}.

Precision electroweak constraints arising from corrections to the
$W$ and $Z$ propagators in these models have previously been
investigated \cite{Csaki:2003zu,Nomura:2003du,Barbieri:2003pr,Davoudiasl:2003me,Burdman:2003ya,Cacciapaglia:2004jz,Davoudiasl:2004pw,Barbieri:2004qk}
 in the continuum and, in the case of weak coupling, using deconstruction
 \cite{Foadi:2003xa,Chivukula:2004kg}.
These analyses begin by specifying a bulk gauge symmetry, ``brane'' kinetic
energy terms for the gauge fields, 
5-D geometry, and fermion couplings. In our analysis, generalizing \cite{Chivukula:2004kg}, we leave the gauge couplings and $f$-constants
of the deconstructed model arbitrary and parameterize the electroweak corrections
in terms of the masses of the heavy vector bosons. We therefore obtain results which
describe arbitrary 5-D geometry, position-dependent couplings, 
or brane kinetic energy terms. 

As our results are computed in terms of the
spectrum of heavy vector bosons, we can directly estimate
the size of electroweak corrections through the constraints unitarity places on the
masses of the heavy vector bosons.  In contrast with other analyses, ours
can determine when non-oblique corrections are relevant, and we can
constrain the form of the non-oblique contributions. 
Unlike Barbieri {\it et. al.} \cite{Barbieri:2003pr}, we are not restricted to the case 
of spatially independent gauge coupling or the case in which the
brane kinetic energy terms dominate those in the bulk. We concur
with the result of Cacciapaglia {\it et.al.} \cite{Cacciapaglia:2004jz} that small or negative $S$ is possible -- though
only at the expense of negative $T$ consistent with $S-4 \cos^2\theta_W T
> {\cal O}(1)$. 

Although we stress our results as they apply to continuum
5-D models, they apply also far from the continuum limit
when only a few extra vector bosons are present. As such, these
results form a generalization of phenomenological analyses \cite{Chivukula:2003wj} of models of extended  electroweak gauge symmetries \cite{Casalbuoni:1985kq,Casalbuoni:1996qt}  motivated by models of hidden local symmetry \cite{Bando:1985ej,Bando:1985rf,Bando:1988ym,Bando:1988br,Harada:2003jx}.

\section{The Model and Its Relatives}

\EPSFIGURE[t]{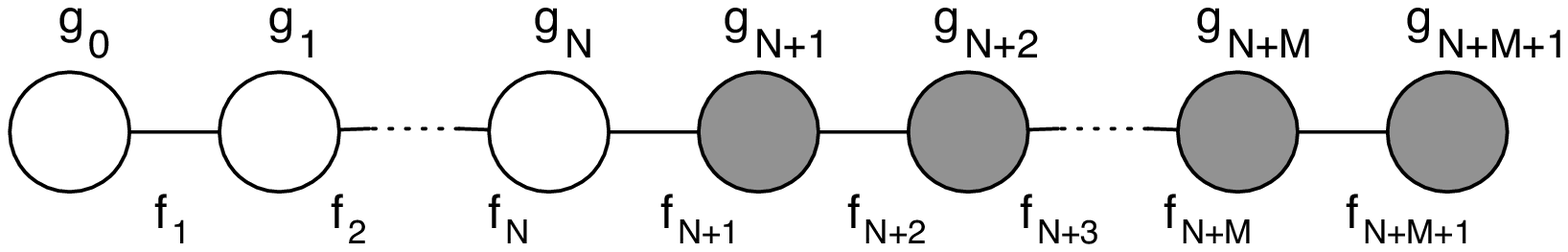,width=0.9\textwidth}
{Moose diagram for the class of models analyzed in this paper. $SU(2)$ gauge groups are shown as open
circles; $U(1)$ gauge groups as shaded circles. The fermions couple to gauge
gauge groups 0 and $N+1$.  The values of the gauge couplings $g_i$ and f-constants $f_i$ are arbitrary.
\label{fig:TheMoose}}

The model we analyze, shown diagrammatically (using ``moose notation'' \cite{Georgi:1986hf,Arkani-Hamed:2001ca}) in Fig. \ref{fig:TheMoose}, incorporates an
$SU(2)^{N+1} \times U(1)^{M+1}$ gauge group, and $N+1$ 
nonlinear $(SU(2)\times SU(2))/SU(2)$ sigma models adjacent to $M$
$(U(1) \times U(1))/U(1)$ sigma models in which the global symmetry groups 
in adjacent sigma models are identified with the corresponding factors of the gauge group.
The Lagrangian for this model at $O(p^2)$ is given by
\begin{equation}
  {\cal L}_2 =
  \frac{1}{4} \sum_{j=1}^{N+M+1} f_j^2 \mbox{tr}\left(
    (D_\mu U_j)^\dagger (D^\mu U_j) \right)
  - \sum_{j=0}^{N+M+1} \dfrac{1}{2g_j^2} \mbox{tr}\left(
    F^j_{\mu\nu} F^{j\mu\nu}
    \right),
\label{lagrangian}
\end{equation}
with
\begin{equation}
  D_\mu U_j = \partial_\mu U_j - i A^{j-1}_\mu U_j 
                               + i U_j A^{j}_\mu,
\end{equation}
where all  gauge fields $A^j_\mu$ $(j=0,1,2,\cdots, N+M+1)$ are dynamical. The first
$N+1$ gauge fields ($j=0,1,\ldots, N$) correspond to $SU(2)$ gauge groups; the other $M+1$ gauge
fields ($j=N+1, N+2, \ldots, N+M+1$) correspond to $U(1)$ gauge groups.  The symmetry breaking between
the $A^{N}_\mu$ and $A^{N+1}_\mu$ follows an $SU(2)_L \times SU(2)_R/SU(2)_V$ symmetry
breaking pattern with the $U(1)$ embedded as the $T_3$-generator of $SU(2)_R$.

The fermions in this model take their weak interactions from the $SU(2)$ group at $j=0$ and their hypercharge interactions from the $U(1)$ group with $j=N+1$, at the interface between the $SU(2)$ and $U(1)$ groups.  The neutral current couplings to the fermions are thus written as
\begin{equation}
J^\mu_3 A^0_\mu + J^\mu_Y A^{N+1}_\mu~,
\label{eq:current}
\end{equation}
while the charged current couplings arise from
\begin{equation}
{1 \over \sqrt{2}} J^\mu_{\pm} A^{0\mp}_\mu~.
\end{equation}

The special case $M=0$ is shown in Fig. \ref{fig:SU2bulk}. Foadi {\it et. al.} \cite{Foadi:2003xa} have made a detailed analysis of a version of the $M=0$ model with a flat geometry (with $f_1 = f_2 = \ldots = f_{N+1}$), constant ``bulk''
coupling ($g_1=g_2=\ldots=g_N$), and weak coupling to fermions
($g_0,\, g_{N+1} \ll g_1$).  They found that, so long as the
model respects unitarity, $S={\cal O}(1)$. This result has since been generalized to arbitrary geometry and
spatially dependent coupling in \cite{Chivukula:2004kg}. The continuum
interpretation of this particular model might seem problematic, as the same fermions
appear to couple at different positions in the discretized fifth dimension.
However, this moose may be redrawn as shown in Fig. \ref{fig:SU2SU2bulk},
and we see that it may equivalently be interpreted in terms of a bulk $SU(2) \times
SU(2)$ gauge group. Motivated by models in a warped background, we
interpret the two ends of the moose as ultraviolet (UV) and infrared (IR) ``branes,''
with the fermions coupled at the UV brane.  In this case, the
 $SU(2) \times SU(2)$ bulk gauge group is broken by boundary conditions to 
a diagonal (custodial) $SU(2)$ at the IR brane, 
and to $SU(2) \times U(1)$ at
the UV brane \cite{Csaki:2003dt,Hebecker:2001jb}.

\EPSFIGURE[t]{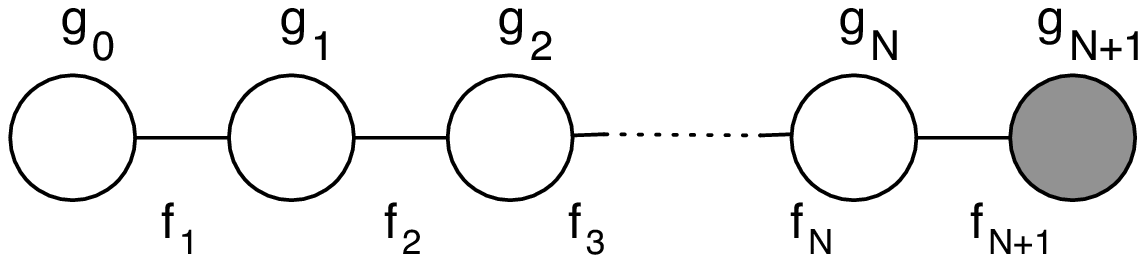,width=0.7\textwidth}
{Moose diagram for the model analyzed in refs. \protect{\cite{Foadi:2003xa,Chivukula:2004kg}}.  This is a special case of the model of Fig. \ref{fig:TheMoose} with $M=0$.
\label{fig:SU2bulk}}

\EPSFIGURE[b]{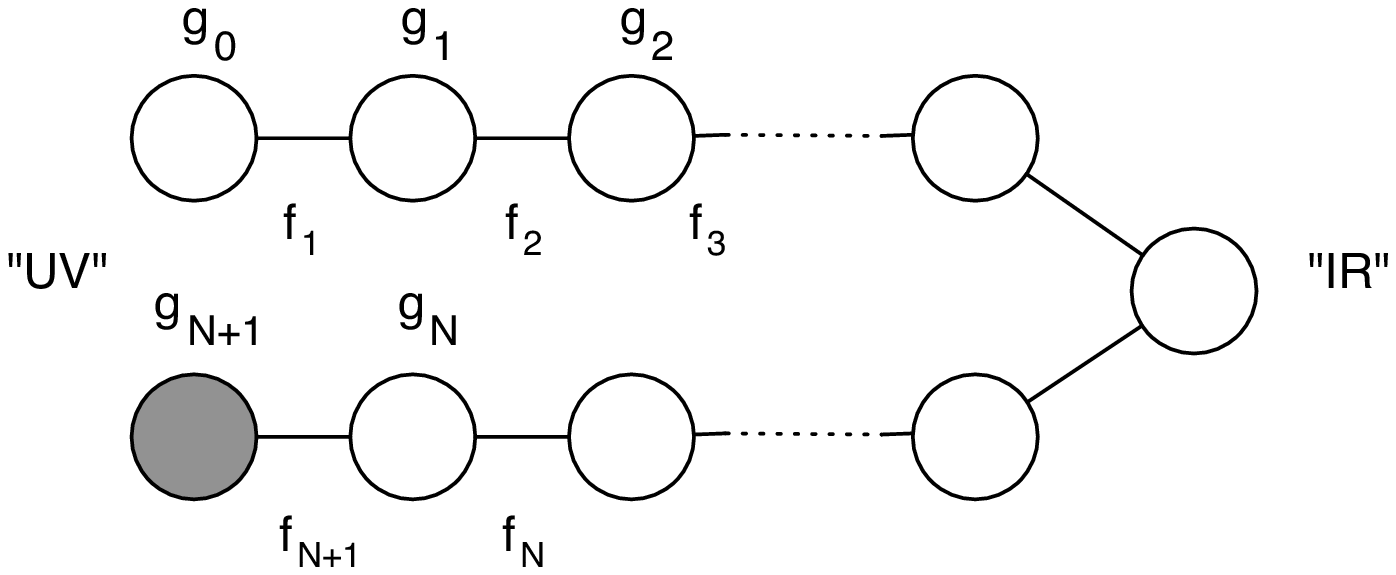,width=0.9\textwidth}
{Moose diagram for a Higgsless model with bulk gauge group $SU(2) \times SU(2)$,
which can be identified with identical to the model illustrated in Fig.
\protect{\ref{fig:SU2bulk}}. Motivated by the warping in an 
AdS space, we label the ``IR'' (infrared) and ``UV'' (ultraviolet) branes.
\label{fig:SU2SU2bulk}}

Inspired by the AdS/CFT correspondence \cite{Maldacena:1998re},
the necessity for an approximate custodial $SU(2)$ symmetry has
motivated the construction of models based on an $SU(2) \times SU(2)
\times U(1)$ gauge group 
\cite{Agashe:2003zs,Csaki:2003zu,Cacciapaglia:2004jz}.
The deconstructed version of the model  discussed in 
\cite{Csaki:2003zu,Cacciapaglia:2004jz} is shown in Fig. \ref{fig:SU2SU2U1bulk}. Here,
the (light) fermions couple on the UV brane, where the boundary conditions break
the bulk $SU(2) \times SU(2) \times U(1)$ gauge group to 
$SU(2) \times U(1)$; at the IR brane, the
boundary conditions break the two $SU(2)$ bulk gauge groups to custodial $SU(2)$.
With an appropriately chosen $U(1)$ IR brane gauge kinetic-energy term,
it has been shown that this model may have a small, or even negative,
$S$ parameter \cite{Cacciapaglia:2004jz}. By ``unfolding'' this moose, we see that
the deconstructed version of this model corresponds to a special case of that shown in 
Fig. \ref{fig:TheMoose}.
An IR $U(1)$  brane kinetic energy term corresponds,
in this language, to having $M \ge 1$ with different $U(1)$ couplings appropriately chosen in the continuum limit.

\EPSFIGURE[t]{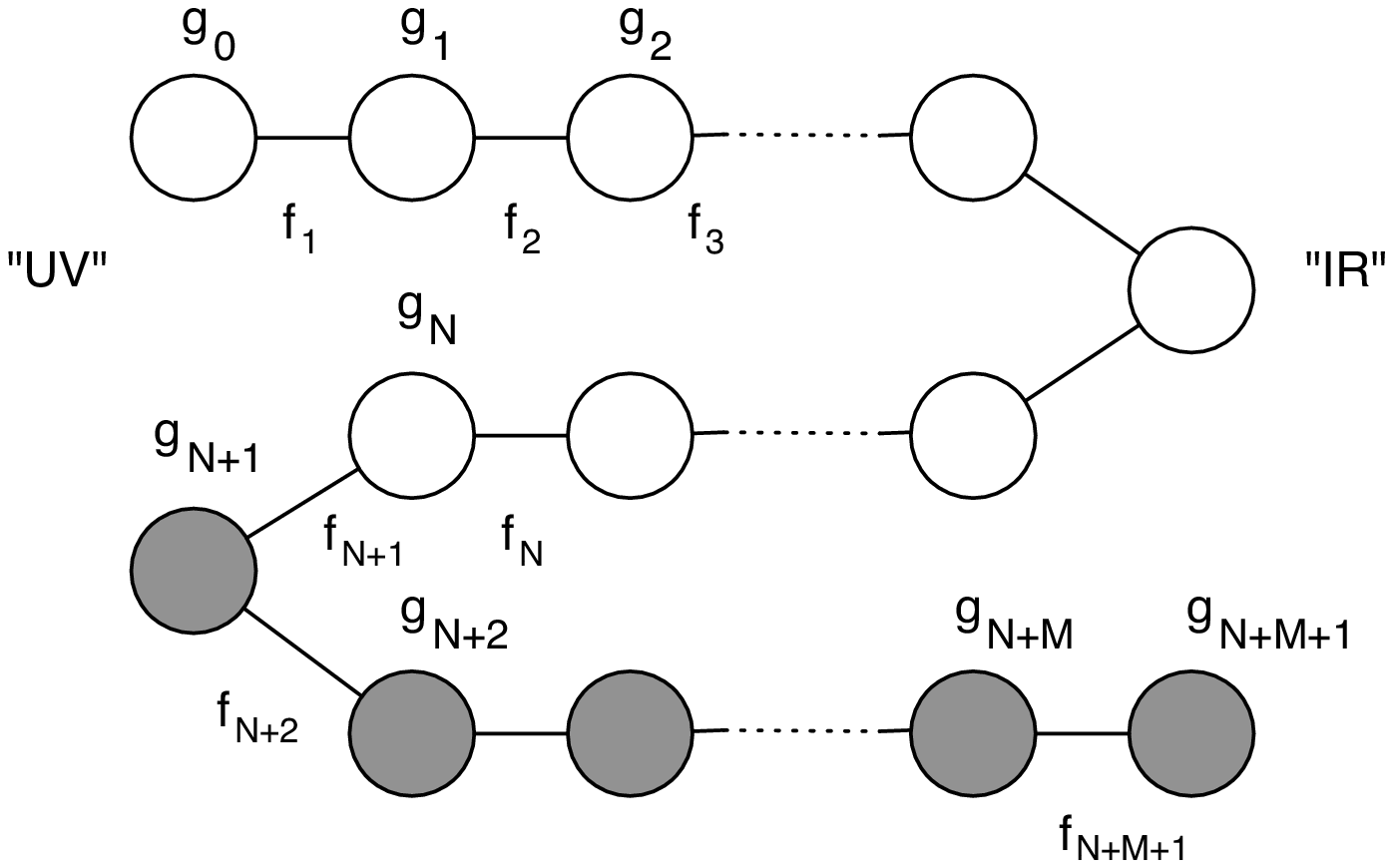,width=0.9\textwidth}
{Moose diagram for the Higgsless model with small $S$ \protect{\cite{Csaki:2003zu,Cacciapaglia:2004jz}}, with bulk gauge group
$SU(2)\times SU(2)\times U(1)$. As proposed the model
has a spatially independent bulk gauge coupling and possible gauge
kinetic energy terms localized at the branes, which corresponds to appropriately
chosen gauge couplings in the continuum limit.
With the (light) fermions coupling on 
the UV brane, this moose is  seen to be a special case of the 
model shown in Fig. 
\protect{\ref{fig:TheMoose}}.
\label{fig:SU2SU2U1bulk}}

\section{The Propagator Matrices}

All four-fermion processes, including those relevant for the electroweak phenomenology of our model, depend on the neutral gauge field
propagator matrix
\begin{equation}
D(Q^2) \equiv \left[ Q^2\, {\cal I} + M^2_{N+M}\right]^{-1}~,
\end{equation}
and the charged propagator matrix
\begin{equation}
\tilde{D}(Q^2) \equiv \left[ Q^2\, {\cal I} + \tilde{M}^2_{N}\right]^{-1}~.
\end{equation}
Here, $M_{N+M}^2$ and $\tilde{M}_N^2$ are, respectively, the mass-squared matrices for the neutral and charged gauge bosons and ${\cal I}$ is the identity matrix.  Consistent with \cite{Chivukula:2004kg}, $Q^2=-q^2$ refers to the
euclidean momentum. 

The neutral vector meson mass-squared matrix is of rank $(N+M+2) \times (N+M+2)$ 
\begin{equation}
{\tiny
M_{N+M}^2 = {1\over 4}
\left(
\begin{array}{c|c|c|c|c|c}
g^2_0 f^2_1& -g_0 g_1 f^2_1 & & &  \\ \hline
-g_0 g_1 f^2_1  & g^2_1(f^2_1+f^2_2) & & &   \\ \hline
 & \ddots & \ddots & \ddots &  \\ \hline
 & & -g^{}_{N} g^{}_{N+1} f^2_{N+1} & g^2_{N+1}(f^2_{N+1} + f^2_{N+2}) & -g^{}_{N+1} g^{}_{N+2} f^2_{N+2} &   \\ \hline
  & & & -g^{}_{N+1} g^{}_{N+2} f^2_{N+2} & g^2_{N+2}(f^2_{N+2}+f^2_{N+3}) & \ddots  \\ \hline
 & & & & \ddots & \ddots \\
\end{array}
\right).
}
\label{eq:neutralmatrix}
\end{equation}
and the charged current vector bosons' mass-squared matrix is the upper $(N+1)  \times (N+1) $ dimensional block of the neutral current $M_{N+M}^2$ matrix
\begin{equation}
{\tiny
\tilde{M}_{N}^2 = {1\over 4}
\left(
\begin{array}{c|c|c|c|c|c}
g^2_0 f^2_1& -g_0 g_1 f^2_1 & & &  \\ \hline
-g_0 g_1 F^2_1  & g^2_1(f^2_1+f^2_2) & -g_1 g_2 f^2_2 & &   \\ \hline
 & -g_1 g_2 f_2^2 & g^2_2(f^2_2+f^2_3) & -g_2 g_3 f^2_3 &  \\ \hline
 & & \ddots & \ddots & \ddots &   \\ \hline
  & & & -g^{}_{N-2} g^{}_{N-1} f^2_{N-1} & g^2_{N-1}(f^2_{N-1}+f^2_{N}) & -g^{}_{N-1} g^{}_{N} f^2_{N} \\ \hline
 & & & & -g^{}_{N-1} g^{}_{N} f^2_{N} & g^2_{N}(f^2_{N}+f^2_{N+1})\\
\end{array}
\right).
}
\label{chargedmatrix}
\end{equation}

It is convenient to define an additional $M\times M$ mass-squared matrix  
${\cal M}_M$,
\begin{eqnarray}
  \lefteqn{
      {\cal M}_M^2 = 
  } \nonumber\\
 & & {\tiny
{1\over 4}
\left(
\begin{array}{c|c|c|c|c|c}
g_{N+2}^2 (f_{N+2}^2+f_{N+3}^2)   & -g^{}_{N+2} g^{}_{N+3} f_{N+3}^2      &
   &  &  & \\ \hline
-g^{}_{N+2} g^{}_{N+3} f_{N+3}^2        & g_{N+3}^2 (f_{N+3}^2+f_{N+4}^2) & 
-g^{}_{N+3} g^{}_{N+4} f_{N+4}^2  &  &  & \\ \hline
                                  & -g^{}_{N+3} g^{}_{N+4} f_{N+4}^2      &
 g_{N+4}^2 (f_{N+4}^2+f_{N+5}^2)  & -g^{}_{N+4} g^{}_{N+5} f_{N+5}^2      &
                                  & \\ \hline
 & & \ddots & \ddots & \ddots  &   \\ \hline
 & & & -g^{}_{K-1} g^{}_{K} f_K^2         &
          g_K^2(f_K^2+f_{K+1}^2) &  -g^{}_K g^{}_{K+1} f_{K+1}^2       \\ \hline
 & & & & -g^{}_K g^{}_{K+1} f_{K+1}^2  &   g_{K+1}^2 f^2_{K+1}         \\
\end{array}
\right)~,
}
\nonumber\\
 & &
\label{eq:mass_matrixM}
\end{eqnarray}
(where we define $K \equiv N+M$)
because it allows us to express the neutral-current matrix (\ref{eq:neutralmatrix}) more compactly as
\begin{equation}
{\tiny
  M_{N+M}^2 = \left(
    \begin{array}{cc|c|cc}
      \multicolumn{2}{c|}{{\tilde M_{N}^2}} & & 
      \multicolumn{2}{c}{} 
      \\ 
      \multicolumn{2}{c|}{} & - g^{}_{N} g^{}_{N+1} f_{N+1}^2/4 & 
      \multicolumn{2}{c}{} 
      \\
      \hline
      \phantom{wwwwww}
      & - g^{}_{N} g^{}_{N+1} f_{N+1}^2/4
      &   g_{N+1}^2 (f_{N+1}^2 + f_{N+2}^2)/4
      & - g^{}_{N+1} g^{}_{N+2} f_{N+2}^2/4
      & 
      \phantom{wwwwww}\\
      \hline
      \multicolumn{2}{c|}{} & - g^{}_{N+1} g^{}_{N+2} f_{N+2}^2/4 & &
      \\
      \multicolumn{2}{c|}{} & & 
      \multicolumn{2}{c}{{\cal M}_{M}^2}
    \end{array}
  \right).
}
\label{eq:mass_matrixN2}
\end{equation}
This form will be especially useful as we proceed.
We also define the propagator matrix 
\begin{equation}
  {\cal D}(Q^2) \equiv \left[ Q^2 {\cal I} + {\cal M}^2_M\right]^{-1}.
\end{equation}
for later use.

Recalling that fermions are charged under only a single SU(2) gauge group (at $j=0$) and a single U(1) group (at $j=N+1$), neutral current four-fermion processes may be derived from the Lagrangian
\begin{equation}
{\cal L}_{nc} = - {1\over 2} g^2_0\, D_{0,0}(Q^2) J^\mu_3 J_{3\mu}
- g_0 g_{N+1}\, D_{0,N+1}(Q^2) J^\mu_3 J_{Y\mu}
-  {1\over 2} g^2_{N+1}\, D_{N+1,N+1}(Q^2) J^\mu_Y J_{Y\mu}~,
\label{nclagrangian}
\end{equation}
and charged-current process from
\begin{equation}
{\cal L}_{cc} = - {1\over 2} g^2_0\, \tilde{D}_{0,0}(Q^2) J^\mu_+ J_{-\mu}~.
\label{cclagrangian}
\end{equation}
where $D_{i,j}$ is the (i,j) element of the gauge field propagator matrix.
From direct inversion of the mass matrix (or see, e.g., \cite{Chivukula:2004kg}) we find that
\begin{equation}
2 \sqrt{2}\, G_F \equiv {1\over 2 } g^2_0\, \tilde{D}_{0,0}(Q^2=0)  = \sum^{N+1}_{j=1} {2 \over f^2_j}  \equiv {2\over v^2}~
\end{equation}
where $v \approx\, 246\, {\rm GeV}$.

We should also consider the nature of the mass eigenstates of this model.  
The neutral vector boson mass matrix (\ref{eq:neutralmatrix}) 
is of a familiar form that has a vanishing determinant, due to a zero eigenvalue.
Physically, this corresponds to a massless neutral gauge field -- the photon.
The lowest massive eigenstates of $M^2_{N+M}$ and $\tilde{M}^2_{N}$ are, respectively, identified as the $Z$ and $W$ bosons. The heavier eigenstates of $M^2_{N+M}$
are labeled by $Z_{\hat{\ell}}$ ($\hat \ell=1,2,\cdots, N+M$), while
those of $\tilde{M}^2_N$ are labeled by $W_{\hat{n}}$ ($\hat n=1,2,\cdots, N$). 
It is also convenient to introduce the eigenvalues of ${\cal M}^2_M$,
which are labeled by ${\cal M}^2_{\hat{m}}$ ($\hat{m}=1,2,\cdots, M$). 

Note that, at tree level, the model specified by the Lagrangian in eqn. (\ref{lagrangian})
is determined by $2N+2M+3$ parameters: $N+M+2$ couplings and $N+M+1$ $f$-constants.
In most of this paper we will specify the values of observable quantities in terms of the
$2N+2M+3$ parameters $e^2$, $M_Z$, $M_{Z\hat{\ell}}$, $M_W$, $M_{W\hat{n}}$, and
${\cal M}_{\hat{m}}$.  In section 6, to facilitate comparison with experiment, we will exchange $M_W$ for the better-measured $G_F$ as an experimental input.  

The propagator matrix elements may be written in a spectral decomposition in terms of the mass eigenstates as follows:
\begin{eqnarray}
  g^2_0 D_{0,0}(Q^2)\equiv [G_{\rm NC}(Q^2)]_{WW} &=& 
    \dfrac{[\xi_\gamma]_{WW}}{Q^2}
   +\dfrac{[\xi_Z]_{WW}}{Q^2 + M_Z^2}
   +\sum_{\hat{k}=1}^{K} \dfrac{[\xi_{Z\hat{k}}]_{WW}}
                                  {Q^2 + M_{Z\hat{k}}^2},
\label{eq:NC_WW}  
  \\
  g_0 g_{N+1} D_{0,N+1}(Q^2) \equiv [G_{\rm NC}(Q^2)]_{WY} &=& 
    \dfrac{[\xi_\gamma]_{WY}}{Q^2}
   +\dfrac{[\xi_Z]_{WY}}{Q^2 + M_Z^2}
   +\sum_{\hat{k}=1}^{K} \dfrac{[\xi_{Z\hat{k}}]_{WY}}
                                  {Q^2 + M_{Z\hat{k}}^2},
\label{eq:NC_WY}  
  \\
  g^2_{N+1} D_{N+1,N+1}(Q^2)\equiv [G_{\rm NC}(Q^2)]_{YY} &=& 
    \dfrac{[\xi_\gamma]_{YY}}{Q^2}
   +\dfrac{[\xi_Z]_{YY}}{Q^2 + M_Z^2}
   +\sum_{\hat{k}=1}^{K} \dfrac{[\xi_{Z\hat{k}}]_{YY}}
                                  {Q^2 + M_{Z\hat{k}}^2},
\label{eq:NC_YY}  
  \\
  g^2_0 \tilde{D}_{0,0}(Q^2)\equiv [G_{\rm CC}(Q^2)]_{WW} &=& 
   \dfrac{[\xi_W]_{WW}}{Q^2 + M_W^2}
   +\sum_{\hat{n}=1}^{N} \dfrac{[\xi_{W\hat{n}}]_{WW}}
                                  {Q^2 + M_{W\hat{n}}^2},
\label{eq:CC_WW}  
\end{eqnarray}
All poles should be simple (i.e. there should be no degenerate mass eigenvalues) because, in the continuum limit, we are analyzing a self-adjoint operator on a finite interval.

Since the neutral bosons couple to only two currents, $J^\mu_3$ and $J^\mu_Y$, 
the three sets of residues in equations (\ref{eq:NC_WW})--(\ref{eq:NC_YY}) must be related. 
Specifically, they satisfy the $N+M+1$ consistency conditions,
\begin{equation}
  [\xi_Z]_{WW} [\xi_Z]_{YY}
  = \left([\xi_Z]_{WY}\right)^2, \qquad
  [\xi_{Z\hat{\ell}}]_{WW} [\xi_{Z\hat{\ell}}]_{YY}
  = \left([\xi_{Z\hat{\ell}}]_{WY}\right)^2 .
\label{consistency}
\end{equation}
In the case of the photon, charge universality further implies
\begin{equation}
  e^2 = [\xi_\gamma]_{WW} = [\xi_\gamma]_{WY} = [\xi_\gamma]_{YY}.
\label{eq:universality}
\end{equation}
We will find expressions for the other residues in the next sections of the paper.

As noted in the introduction, a number of models in the literature correspond to the special case of this theory where there is only a single $U(1)$ group.  In our notation, this corresponds to $M =0$.  Equation (\ref{eq:mass_matrixN2}) for the neutral mass-squared matrix simplifies to
\begin{equation}
{\tiny
  M_{N+M}^2 = \left(
    \begin{array}{cc|c}
      \multicolumn{2}{c|}{{\tilde M_{N}^2}} & 
      \\ 
      \multicolumn{2}{c|}{} & - g^{}_{N} g^{}_{N+1} f_{N+1}^2 
      \\
      \hline
      \phantom{wwwwww}
      - g^{}_{N} g^{}_{N+1} f_{N+1}^2 
      & & g_{N+1}^2 f_{N+1}^2 
    \end{array}
  \right).
}
\end{equation}
The ${\cal{M}}_M^2$ matrix does not arise, and all observables can be specified in
terms of the $2N +3$ quantities  $e^2$, $M_Z$, $M_{Z\hat{\ell}}$, $M_W$, and $M_{W\hat{n}}$.
Hence, in the results that follow, the case of a single $U(1)$ group will correspond
to taking $M=0$ and setting multiplicative factors involving the ${\cal{M}}_{\hat{m}}$ to unity.

\section{Correlation Functions and Couplings}

We now find the values of the coefficients $\xi$ in the spectral representation of the neutral and charged-current correlation functions (\ref{eq:NC_WW})-(\ref{eq:CC_WW}).  These directly show the contributions of the various weak bosons to four-fermion processes.  We can also write the couplings of the $Z$ and $W$ bosons in terms of the residues
\begin{equation}
 ( \sqrt{[\xi_Z]_{WW}} \, J_{3\mu} - \sqrt{[\xi_Z]_{YY}}\,  J_{Y\mu} )\, Z^\mu
 \label{eq:zzcoup}
\end{equation}
\begin{equation}
 \sqrt{ \dfrac{[\xi_W]_{WW}}{2}} \, J^\mp_\mu \,W^{\pm\mu}
\end{equation}
and see how the existence of the heavy gauge bosons alters these couplings from their standard model values.  We will find that the $W$ and $Z$ couplings approach their tree-level standard model values  in the limit $M_{Z\hat{k}}, M_{W\hat{n}}, {\cal M}_{\hat{m}} \to  \infty$.

\subsection{\protect{$[G_{\rm NC}(Q^2)]_{YY}$ and its residues}}

We will start by determining the pure hypercharge element of the propagator matrix,
$D(Q^2)_{N+1,N+1}$.  This will lead us to the residues $[\xi_Z]_{YY}$ and $[\xi_{Z\hat{\ell}}]_{YY}$ of $[G_{\rm NC}(Q^2)]_{YY}=g_{N+1}^2 D(Q^2)_{N+1,N+1}$, which reveal the relative contributions of $Z$ and $Z_{\hat{\ell}}$ exchange to four-fermion processes in the hypercharge channel.   

Direct calculation of the matrix inverse involved in $D_{N+1,N+1}(Q^2)$ involves
the computation of the cofactor related to the $(N+1,N+1)$ element of the
matrix [$Q^2 {\cal I}_{N+M+2} +M^2_{N+M}$]. Inspection of the neutral-boson mass matrix eqn. (\ref{eq:mass_matrixN2}) shows that this cofactor is given by 
\begin{equation}
  \det[\tilde D^{-1}(Q^2)] \det[{\cal D}^{-1}(Q^2)].
\end{equation}
We therefore obtain
\begin{eqnarray}
  D(Q^2)_{N+1,N+1} 
  &=& 
  \dfrac{\det[\tilde D^{-1}(Q^2)] \det[{\cal D}^{-1}(Q^2)]}
        {\det[D^{-1}(Q^2)]}
  \nonumber\\
  &=&
  \dfrac{(Q^2 + M_W^2)
         \left(\prod_{\hat{n}=1}^N (Q^2 + M_{W\hat{n}}^2)\right)
         \left(\prod_{\hat{m}=1}^M (Q^2 + {\cal M}_{\hat{m}}^2)\right)
        }
        {Q^2 (Q^2 + M_Z^2)
         \prod_{\hat{k}=1}^K (Q^2 + M_{Z\hat{k}}^2)
        }~,
\label{eq:yy_26}
\end{eqnarray}
by writing the determinants in terms of the eigenvalues.

As noted earlier, charge universality for the photon tells us that the residue of $[G_{\rm NC}(Q^2)]_{YY}$ at $Q^2=0$ is $e^2$.  Using equations (\ref{eq:NC_YY}) and (\ref{eq:yy_26}) we find
\begin{equation}
  e^2 = g_{N+1}^2 
       \dfrac{M_W^2 \left(\prod_{\hat{n}=1}^{N} M_{W\hat{n}}^2\right)
              \left(\prod_{\hat{m}=1}^{M} {\cal M}_{\hat{m}}^2\right)}
             {M_Z^2 \prod_{\hat{k}=1}^{K} M_{Z\hat{k}}^2},
\end{equation}
and therefore
\begin{equation}
  [G_{\rm NC}(Q^2)]_{YY}
  = \dfrac{e^2}{Q^2}\dfrac{[Q^2+M_W^2]M_Z^2}{M_W^2[Q^2+M_Z^2]}
    \left(
      \prod_{\hat{k}=1}^{K}
      \dfrac{M_{Z\hat{k}}^2}{Q^2+M_{Z\hat{k}}^2}
    \right)
    \left(
      \prod_{\hat{n}=1}^{N}
      \dfrac{Q^2+M_{W\hat{n}}^2}{M_{W\hat{n}}^2}
    \right)
    \left(
      \prod_{\hat{m}=1}^{M}
      \dfrac{Q^2+{\cal M}_{\hat{m}}^2}{{\cal M}_{\hat{m}}^2}
    \right).
\label{eq:exact_YY}
\end{equation}

Reading off the residues of the poles in Eq.(\ref{eq:exact_YY}), we obtain
\begin{eqnarray}
  {}
  [\xi_Z]_{YY} 
  &=& e^2 \dfrac{M_Z^2- M_W^2}{M_W^2}
      \left(\prod_{\hat{k}=1}^K
      \dfrac{M_{Z\hat{k}}^2}{M_{Z\hat{k}}^2-M_Z^2}
      \right)
      \left(\prod_{\hat{n}=1}^N
      \dfrac{M_{W\hat{n}}^2-M_Z^2}{M_{W\hat{n}}^2}
      \right) 
  \nonumber\\
  & & \qquad \times 
      \left(\prod_{\hat{m}=1}^M
      \dfrac{{\cal M}_{\hat{m}}^2-M_Z^2}{{\cal M}_{\hat{m}}^2}
      \right),
\label{eq:Z_residue_YY}
 \\
  {}
  [\xi_{Z\hat{\ell}}]_{YY} &=&
   - e^2 \dfrac{[M_{Z\hat{\ell}}^2 - M_W^2]M_Z^2}
             {M_W^2 [M_{Z\hat{\ell}}^2 - M_Z^2]}
      \left(\prod_{\hat{k} \ne \hat{\ell}}
      \dfrac{M_{Z\hat{k}}^2}{M_{Z\hat{k}}^2-M_{Z\hat{\ell}}^2}
      \right)
      \left(\prod_{\hat{n}=1}^N
      \dfrac{M_{W\hat{n}}^2-M_{Z\hat{\ell}}^2}{M_{W\hat{n}}^2}
      \right)
  \nonumber\\
  & & \qquad \times 
      \left(\prod_{\hat{m}=1}^M
      \dfrac{{\cal M}_{\hat{m}}^2-M_{Z\hat{\ell}}^2}{{\cal M}_{\hat{m}}^2}
      \right).
\label{eq:Z_ell_residue_YY}
\end{eqnarray}
Note that, in the limit $M_{Z\hat{k}}, M_{W\hat{n}}, {\cal M}_{\hat{m}} \to 
\infty$, $[\xi_Z]_{YY} \to e^2 (M^2_Z - M^2_W)/M^2_W$, so that the correct standard
model tree-level value of the $Z$-boson coupling to hypercharge is recovered.

\subsection{\protect{$[G_{\rm NC}(Q^2)]_{WY}$ and its residues}}

Next, we turn to the weak-hypercharge interference term in equation (\ref{nclagrangian}).
Direct calculation of the matrix inverse involved in $D_{0,N+1}(Q^2)$ involves
the computation of the cofactor related to the $(0,N+1)$ element of the
matrix $Q^2 {\cal I} +M^2_{N+M}$.  Inspection of the neutral-boson mass matrix
eqn. (\ref{eq:neutralmatrix}) shows that the cofactor is the determinant of a matrix whose
upper $(N+1) \times (N+1)$ block is upper-diagonal, with diagonal entries $\{-g_0 g_1 F^2_1/4, -g_1 g_2 F^2_2/4, \ldots, -g_N g_{N+1} F^2_{N+1}/4 \}$ that are independent of $Q^2$; its determinant is therefore also constant. The lower $M \times M$ block of the cofactor is just ${\cal M}^2_M$.  Therefore
\begin{equation}
  [G_{\rm NC}(Q^2)]_{WY} = C
  \dfrac{\det[{\cal D}^{-1}(Q^2)]}{\det[D^{-1}(Q^2)]}
  = C
  \dfrac{\prod_{\hat{m}=1}^{M} (Q^2 + {\cal M}_{\hat{m}}^2)}
        {Q^2 (Q^2 + M_Z^2) 
         \prod_{\hat{k}=1}^{K} (Q^2 + M_{Z\hat{k}}^2)}.
\end{equation}
where $C$ is a constant.

Requiring the residue of the photon pole at  $Q^2=0$ to equal $e^2$ determines the value of $C$ to be $[e^2 M_Z^2 \prod_{\hat{k}=1}^{K} M_{Z\hat{k}}^2]  /  [\prod_{\hat{m}=1}^{M} {\cal{M}}_{\hat{m}}^2 ]$.  
We thus obtain
\begin{equation}
  [G_{\rm NC}(Q^2)]_{WY} = 
  \dfrac{e^2 M_Z^2}{Q^2[Q^2 + M_Z^2]}
        \left(\prod_{\hat{k}=1}^{K} 
        \dfrac{M_{Z \hat{k}}^2}{Q^2 + M_{Z \hat{k}}^2}\right)
        \left(\prod_{\hat{m}=1}^{M} 
        \dfrac{Q^2 + {\cal M}_{\hat{m}}^2}{{\cal M}_{\hat{m}}^2}
        \right).
\label{eq:exact_WY}
\end{equation}

Reading off the residues of the poles in Eq.(\ref{eq:exact_WY}), we find
\begin{eqnarray}
  {}
  [\xi_Z]_{WY} &=&
   - e^2 \left(\prod_{\hat{k}=1}^{K} 
         \dfrac{M_{Z\hat{k}}^2}
               {M_{Z\hat{k}}^2 - M_Z^2}
    \right)
    \left(\prod_{\hat{m}=1}^{M} 
         \dfrac{{\cal M}_{\hat{m}}^2 - M_Z^2}
               {{\cal M}_{\hat{m}}^2}
    \right),
\label{eq:Z_residue_WY}
  \\
  {}
  [\xi_{Z\hat{\ell}}]_{WY} &=&
    e^2 \dfrac{M_Z^2}{M_{Z\hat{\ell}}^2- M_Z^2}
    \left(\prod_{\hat{k}\ne \hat{\ell}}
         \dfrac{M_{Z\hat{k}}^2}
               {M_{Z\hat{k}}^2  - M_{Z\hat{\ell}}^2 }
    \right)
    \left(\prod_{\hat{m}=1}^{M} 
         \dfrac{{\cal M}_{\hat{m}}^2 - M_{Z\hat{\ell}}^2}
               {{\cal M}_{\hat{m}}^2}
    \right).
\label{eq:Z_ell_residue_WY}
\end{eqnarray}
Note that, in the limit $M_{Z\hat{k}},\, {\cal M}_{\hat{m}} \to \infty$, $[\xi_{Z}]_{WY}
\to -e^2$ as at tree-level in the standard model.

\subsection{Consistency Conditions and \protect{$[G_{NC}(Q^2)]_{WW}$}}

Having determined the YY and WY residues for the neutral boson poles, we can use the consistency conditions of equation (\ref{consistency}) to deduce the WW residues as well
  \begin{eqnarray}
  {}
  [\xi_Z]_{WW} &=&
   \dfrac{e^2 M_W^2}{M_Z^2 - M_W^2}
   \left(
   \prod_{\hat{k}=1}^{K} 
     \dfrac{M_{Z\hat{k}}^2}
           {M_{Z\hat{k}}^2 - M_Z^2}
   \right)
   \left(
   \prod_{\hat{n}=1}^{N} 
     \dfrac{M_{W\hat{n}}^2}
           {M_{W\hat{n}}^2 - M_Z^2}
   \right)
  \nonumber\\
  & & \qquad \times
   \left(
   \prod_{\hat{m}=1}^{M} 
     \dfrac{{\cal M}_{\hat{m}}^2 - M_Z^2}{{\cal M}_{\hat{m}}^2}
   \right)~,
\label{eq:Z_residue_WW}
\end{eqnarray}
and
  \begin{eqnarray}
  {}
  [\xi_{Z\hat{\ell}}]_{WW} &=&
-   \dfrac{e^2 M_W^2 M^2_Z}{(M_{Z\hat{\ell}}^2 - M_W^2)(M_{Z\hat{\ell}}^2 - M_Z^2)}
   \left(
   \prod_{\hat{k}\neq \hat{\ell}} 
     \dfrac{M_{Z\hat{k}}^2}
           {M_{Z\hat{k}}^2 - M_{Z\hat{\ell}}^2}
   \right)
   \left(
   \prod_{\hat{n}=1}^{N} 
     \dfrac{M_{W\hat{n}}^2}
           {M_{W\hat{n}}^2 - M_{Z\hat{\ell}}^2}
   \right)
  \nonumber\\
  & & \qquad \times
   \left(
   \prod_{\hat{m}=1}^{M} 
     \dfrac{{\cal M}_{\hat{m}}^2 - M_{Z\hat{\ell}}^2}{{\cal M}_{\hat{m}}^2}
   \right)~.
\label{eq:Z_ell_residue_WW}
\end{eqnarray}
In the limit 
$M_{Z\hat{k}}, M_{W\hat{n}}, {\cal M}_{\hat{m}} \to 
\infty$,  one finds
$[\xi_Z]_{WW} \to e^2 M^2_W/(M^2_Z-M^2_W)$, so that the correct standard
model tree-level form of the $Z$ boson coupling to weak charge is recovered.

\subsection{Custodial Symmetry  and \protect{$[G_{CC}(Q^2)]_{WW}$}}

\EPSFIGURE[t]{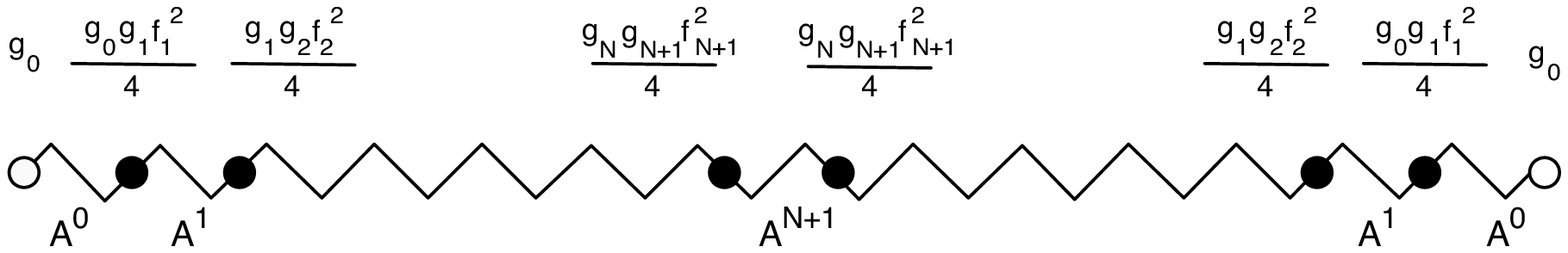,width=0.9\textwidth}
{Leading diagram at high-$Q^2$ ($Q^2\gg M_{W\hat{n}}^2, M_{Z\hat{\ell}}^2$) which 
distinguishes $[G_{NC}(Q^2)]_{WW}$ from
$[G_{CC}(Q^2)]_{WW}$, resulting in the behavior of eqn. (\protect{\ref{eq:high_energy_WW}}).
The open circles at either end represent the weak current $J_3^\mu$ (see eqn. 
(\protect{\ref{eq:current}})), the solid circles represent the off-diagonal vector-boson mass
matrix elements in eqn. (\protect{\ref{eq:neutralmatrix}}),
and the gauge-bosons are represented by the zigzag lines. The difference between
$[G_{NC}(Q^2)]_{WW}$ and $[G_{CC}(Q^2)]_{WW}$ arises from the couplings of the 
$U(1)$ factors, and the leading contribution comes from the factor with coupling $g_{N+1}$.
\label{fig:custodial}}

Finally, we need to determine the charged gauge boson correlation functions.
Here, we have no consistency conditions to guide us, but we 
can exploit custodial symmetry.

The value of the difference of correlation functions
\begin{equation}
  [G_{\rm NC}(Q^2)]_{WW} - [G_{\rm CC}(Q^2)]_{WW}.
\label{eq:NC-CC_WW}
\end{equation}
vanishes in the
limit of exact custodial symmetry, $g_{N+1}\to 0$.
Since the violation of the custodial symmetry $g_{N+1}\ne 0$ is
localized at the $(N+1)$th site, 
Eq.(\ref{eq:NC-CC_WW}) must fall at large $Q^2$ like
\begin{equation}
  [G_{\rm NC}(Q^2)]_{WW} - [G_{\rm CC}(Q^2)]_{WW}
  \propto \dfrac{g_{N+1}^2}{(Q^2)^{2N+3}} ,  
\label{eq:high_energy_WW}
\end{equation}
for $Q^2\gg M_{W\hat{n}}^2, M_{Z\hat{\ell}}^2$, as shown in fig. (\ref{fig:custodial}).

Consider using the spectral decompositions of the correlation functions from 
Eq.(\ref{eq:NC_WW}) and Eq.(\ref{eq:CC_WW})
to evaluate eqn. (\ref{eq:NC-CC_WW}).  Forming a common
denominator, we see that the only way to achieve the high energy behavior of
Eq.(\ref{eq:high_energy_WW}) is if the numerator includes a
polynomial in $Q^2$ of order $M$.   Denoting that polynomial as ${\cal R}$,
we can write
\begin{equation}
  [G_{\rm NC}(Q^2)]_{WW}  - [G_{\rm CC}(Q^2)]_{WW} 
= \dfrac{e^2 M_W^2 M_Z^2 \,{\cal R}(Q^2)}{Q^2[Q^2+M_W^2][Q^2 + M_Z^2]}
        \left(
        \prod_{\hat{k}=1}^{K} 
        \dfrac{M_{Z \hat{k}}^2}
              {Q^2 + M_{Z \hat{k}}^2}
        \right)
        \left(
        \prod_{\hat{n}=1}^{N} 
        \dfrac{M_{W \hat{n}}^2}
              {Q^2 + M_{W \hat{n}}^2}
        \right)\ .
\label{eq:exact_WW1}
\end{equation}
The residues $[\xi_Z]_{WW}$ and $[\xi_{Z\hat{\ell}}]_{WW}$ extracted from
Eq.(\ref{eq:exact_WW1}) must satisfy the $N+M+1$
consistency conditions of eqn. (\ref{consistency}). 
The number of consistency conditions is $N+M+1$, while the polynomial
${\cal R}$ possesses $M$ parameters -- thus the polynomial 
${\cal R}$ is (over) determined solely by the consistency conditions. 

To evaluate ${\cal R}$, we begin by defining a more compact notation
\begin{eqnarray}
   R_Z(Q^2) &\equiv& \dfrac{1}{Q^2}\dfrac{M_Z^2} {Q^2+M_Z^2}
        \prod_{\hat{\ell}=1}^K 
          \dfrac{M_{Z\hat{\ell}}^2}{Q^2+M_{Z\hat{\ell}}^2} ,
   \\
 R_W(Q^2) &\equiv& \dfrac{M_W^2} {Q^2+M_W^2}
        \prod_{\hat{n}=1}^N 
          \dfrac{M_{W\hat{n}}^2}{Q^2+M_{W\hat{n}}^2} ,
   \\
   {\cal R}_m(Q^2) &\equiv&
      \prod_{\hat{m}=1}^M
      \dfrac{{\cal M}_{\hat{m}}^2}{Q^2  + {\cal M}_{\hat{m}}^2}.
\end{eqnarray}
We can rewrite the correlation functions in this language as
\begin{equation}
[G_{NC}(Q^2)]_{YY}  =  {e^2} {R_Z(Q^2) \over R_W(Q^2) {\cal R}_m (Q^2)}~,
\label{eq:gncyy}
\end{equation}\begin{equation}
[G_{NC} (Q^2)]_{WY}  =    {e^2} {R_Z(Q^2) \over {\cal R}_m(Q^2)}~,
\label{eq:gncwy}
\end{equation}
\begin{equation}
[G_{NC}(Q^2)]_{WW} - [G_{CC}(Q^2)]_{WW} =  {e^2} {\cal R}(Q^2)R_Z(Q^2) R_W(Q^2)
\label{eq:gnc-diff}
\end{equation}
where ${\cal R}$ is still our unknown polynomial.

Likewise, the residues can be written rather simply in this notation.  In particular, we can evaluate $[\xi_Z]_{WW}$ as the residue of the Z-boson pole of equation (\ref{eq:gnc-diff}).  Since $R_Z(Q^2)$ includes a simple pole at  $Q^2 = -M_Z^2$,  the unknown polynomial ${\cal R}$ will be analytic at that point.
\begin{equation}
[\xi_Z]_{WW} = e^2 {\cal R}(-M_Z^2) R_W (-M_Z^2) \left[ (Q^2 + M_Z^2) R_Z(Q^2) \right]_{Q^2 = - M_Z^2}
\label{eq:method1}
\end{equation}
Alternatively, we can rewrite in this compact notation the expression (\ref{eq:Z_residue_WW}) for $[\xi_Z]_{WW}$ which we derived earlier from 
$[\xi_Z]_{YY}$ and $[\xi_Z]_{WY}$  and the consistency conditions  (\ref{consistency}) 
\begin{equation}
[\xi_Z]_{WW} = e^2 \dfrac{ R_W (-M_Z^2) \left[ (Q^2 + M_Z^2) R_Z(Q^2) \right]_{Q^2 = - M_Z^2}}
{ {\cal R}_m(-M_Z^2)}
\label{eq:method2}
\end{equation}
Comparing expressions (\ref{eq:method1}) and (\ref{eq:method2}) for $[\xi_Z]_{WW}$ and making similar comparisons for the two ways of finding the residues of the $Z_{\hat{\ell}}$ poles shows that
\begin{equation}
  {\cal R}(Q^2) =
      \prod_{\hat{m}=1}^M
      \dfrac{Q^2  + {\cal M}_{\hat{m}}^2}{{\cal M}_{\hat{m}}^2} = \dfrac{1}{{\cal{R}}_m(Q^2)}.
\end{equation}
is the polynomial we seek.  

As a result, we find
\begin{equation}
  [G_{\rm NC}(Q^2)]_{WW}  - [G_{\rm CC}(Q^2)]_{WW} 
  =  {e^2} \dfrac{ R_Z(Q^2) R_W(Q^2)} {{\cal R}_m(Q^2)}
\end{equation}
with the interesting corollary from comparing this with equations (\ref{eq:gncyy}) and (\ref{eq:gncwy})
\begin{equation}
[G_{NC}(Q^2)]_{WY}^2 = [G_{NC}(Q^2)]_{YY} \cdot \left( 
[G_{NC}(Q^2)]_{WW} - [G_{CC}(Q^2)]_{WW}
\right)~.
\label{eq:relationship}
\end{equation}
Note that the $W$ poles in $[G_{CC}(Q^2)]_{WW}$ are cancelled by the corresponding
zeros in $[G_{NC}(Q^2)]_{YY}$. 

Finally, we can write the difference of the neutral and charged current correlation functions in our original notation as
\begin{eqnarray}
\lefteqn{  
  [G_{\rm NC}(Q^2)]_{WW}  - [G_{\rm CC}(Q^2)]_{WW} 
} \nonumber\\
  &=& 
        \dfrac{e^2 M_W^2 M_Z^2}{Q^2[Q^2+M_W^2][Q^2 + M_Z^2]}
        \left(
        \prod_{\hat{k}=1}^{K} 
        \dfrac{M_{Z \hat{k}}^2}
              {Q^2 + M_{Z \hat{k}}^2}
        \right)
        \left(
        \prod_{\hat{n}=1}^{N} 
        \dfrac{M_{W \hat{n}}^2}
              {Q^2 + M_{W \hat{n}}^2}
        \right)
  \nonumber\\
  & & \qquad \times
        \left(
          \prod_{\hat{m}=1}^M
          \dfrac{Q^2  + {\cal M}_{\hat{m}}^2}{{\cal M}_{\hat{m}}^2}
        \right).
\label{eq:exact_WW}
\end{eqnarray}
Reading the residues of the poles from Eq.(\ref{eq:exact_WW}), we obtain the same
results as before (equations (\ref{eq:Z_residue_WW}) and (\ref{eq:Z_ell_residue_WW})) for the $Z$ and $Z_{\hat{\ell}}$ residues.  We also obtain the values of the last residues we need, those for the $W$ and $W_{\hat{\ell}}$ poles
\begin{equation}
  {}
  [\xi_W]_{WW} =
   \dfrac{e^2 M_Z^2}{M_Z^2 - M_W^2}
   \left(
   \prod_{\hat{k}=1}^{K} 
     \dfrac{M_{Z\hat{k}}^2}
           {M_{Z\hat{k}}^2 - M_W^2}
   \right)
   \left(
   \prod_{\hat{n}=1}^{N} 
     \dfrac{M_{W\hat{n}}^2}
           {M_{W\hat{n}}^2 - M_W^2}
   \right)
   \left(
   \prod_{\hat{m}=1}^{M} 
     \dfrac{{\cal M}_{\hat{m}}^2 - M_W^2}{{\cal M}_{\hat{m}}^2}
   \right),
\label{eq:W_residue_WW}
\end{equation}
\begin{eqnarray}
  {}
  [\xi_{W\hat{\ell}}]_{WW} &=&
    \dfrac{e^2 M^2_W M_Z^2}{(M_{W\hat{\ell}}^2 - M_W^2)(M_{W\hat{\ell}}^2 - M_Z^2)}
   \left(
   \prod_{\hat{k}=1}^K
     \dfrac{M_{Z\hat{k}}^2}
           {M_{Z\hat{k}}^2 - M_{W\hat{\ell}}^2}
   \right)
   \left(
   \prod_{\hat{n}\neq \hat{\ell}} 
     \dfrac{M_{W\hat{n}}^2}
           {M_{W\hat{n}}^2 - M_{W\hat{\ell}}^2}
   \right)
  \nonumber\\
  & & \qquad \times
   \left(
   \prod_{\hat{m}=1}^{M} 
     \dfrac{{\cal M}_{\hat{m}}^2 - M_{W\hat{\ell}}^2}{{\cal M}_{\hat{m}}^2}
   \right).
\label{eq:W_ell_residue_WW}
  \end{eqnarray}
In the limit 
$M_{Z\hat{k}}, M_{W\hat{n}}, {\cal M}_{\hat{m}} \to 
\infty$, we have $[\xi_W]_{WW} \to e^2 M^2_Z/(M^2_Z - M^2_W)$ and the correct standard
model tree-level value of the $W$-boson coupling to weak charge is recovered.

Comparison of equations (\ref{eq:Z_residue_WW}) with equation (\ref{eq:W_residue_WW}) 
shows that the formulae for $[\xi_Z]_{WW}$ and  $[\xi_W]_{WW}$ are related by the simple 
exchange $[Z \leftrightarrow W]$.  A similar relationship holds for  $[\xi_{W\hat{\ell}}]_{WW}$ and 
$[\xi_{Z\hat{\ell}}]_{WW}$.   While this relationship seems intuitive in hindsight, a first-principles 
proof starting from the relationship between $M_{N+M}^2$ and $\tilde{M}_{N}^2$ has not been found.

\section{The Low-Energy $\rho$ Parameter}

In discussing low-energy interactions it is conventional to rewrite  the neutral-current Lagrangian of eqn. (\ref{nclagrangian}) in terms of weak and electromagnetic currents
as
\begin{equation}
{\cal L}_{nc} = - {1\over 2} A(Q^2)  J^\mu_3 J_{3\mu} - B(Q^2) J^\mu_3 J_{Q\mu}
- {1\over 2} C(Q^2) J^\mu_Q J_{Q\mu}~,
\end{equation}
where
\begin{eqnarray}
A(Q^2) & = & [G_{NC}(Q^2)]_{WW} - 2 [G_{NC}(Q^2)]_{WY}+ [G_{NC}(Q^2)]_{YY} \\
B(Q^2) & = & [G_{NC}(Q^2)]_{WY} - [G_{NC}(Q^2)]_{YY} \\
C(Q^2) & = & [G_{NC}(Q^2)]_{YY}~.
\end{eqnarray}
As required, the photon pole cancels in the expressions for $A(Q^2)$ and
$B(Q^2)$.
Charged-current interactions will continue to be expressed by eqn. (\ref{cclagrangian})
\begin{equation}
{\cal L}_{cc} = - {1\over 2} [G_{CC}(Q^2)]_{WW}  J^\mu_+ J_{-\mu}~.
\label{cclagrangiani}
\end{equation}

It is informative to compare the overall strength of the low-energy neutral current interactions
with that of the charged-current interactions. In assessing the strength of low-energy neutral
currents, we must  calculate $A(Q^2=0)$. This is conveniently done by evaluating the derivative
\begin{equation}
\left({d\{ Q^2 A(Q^2)\}  \over dQ^2}\right)_{Q^2=0}~.
\label{eq:deriv-form}
\end{equation}
Similarly, since the charged-current propagator has no pole at $Q^2=0$,
one finds 
\begin{equation}
\left({d\{ Q^2 [G_{CC}(Q^2)]_{WW}\} \over dQ^2}\right)_{Q^2=0}
= [G_{CC}(0)]_{WW}~.
\end{equation}
Using eqn. (\ref{eq:relationship}) and the fact that the products $R_W(Q^2)$ and ${\cal R}_m(Q^2)$ are well-behaved at $Q^2=0$, with 
${\cal R}_m(0) =  R_W(0)  = 1$, one finds that expression (\ref{eq:deriv-form}) is equal to $[G_{CC}(0)]_{WW}$.  The low-energy $\rho$ parameter is identically 1:
\begin{equation}
\rho \equiv {1 \over [G_{CC}(0)]_{WW}} 
\left({d\{ Q^2 A(Q^2)\}  \over dQ^2}\right)_{Q^2=0} = 1~.
\label{eq:rho-parameter}
\end{equation}
Therefore, in the sense defined here, the strengths of the charged-current and weak neutral-current interactions at low energy are the same.\footnote{Note that the strength of low-energy
neutral and charged currents can be the same, while the ratio of $W$ and $Z$ masses can differ from
the standard model value, if the contributions from heavy boson
exchange are significant. This is the case in  the model discussed in ref. 
\protect{\cite{Csaki:2003dt}}, for example.} 
This is an extension of the result in  \cite{Chivukula:2003wj} and is due to our specifying that the fermion hypercharges arise from the $U(1)$ group with $j=N+1$.

Finally, we note for future use that the phenomenologically relevant low-energy neutral current coupling for atomic parity violation (since Z $\approx$ N for most nuclei) is $B(Q^2)$,
 which couples $J^\mu_3$ to the electromagnetic current $J_{Q\mu}$. 
In this quantity, the photon pole cancels and
we find that
\begin{equation}
B(Q^2=0) \equiv \left({d\{ Q^2 B(Q^2)\}  \over dQ^2}\right)_{Q^2=0}
= -\,{e^2 \over M^2_W} 
\left[1+\sum^N_{\hat{l}=1} {M^2_W \over M^2_{W\hat{l}}} \right]~.
\label{eq:APV}
\end{equation}
We will find in the next section that in some limits of the model $B(Q^2=0)$ arises from
Z-boson exchange alone, while in others the exchange of heavy neutral vector bosons must be included at low energies to give the correct value.

\section{Electroweak Parameters}

From the calculation of the propagator and corresponding residues, we see that the
size of the couplings of the heavy resonances to fermions depends crucially on the
isospin asymmetry of these states. In the isospin symmetric limit, $g_{N+1} \to 0$,
$M_Z \to M_W$ and $M_{Z\hat{k}} \to \{ M_{W\hat{n}}\ {\rm or}\  {\cal M}_{\hat{m}}\}$. The couplings
of the heavy resonances depend on the size of $\Delta M^2_{Z\hat{k}}$, the mass splitting between $M^2_{Z\hat{k}}$
and the corresponding mass $M^2_{W\hat{n}}$ or ${\cal M}^2_{\hat{m}}$ with which it becomes
paired in the isospin symmetric limit.  In the subsections below we examine limits with different values of the mass splitting $\Delta M^2_{Z\hat{k}}$.

Note that up to this point we have used $M_W$ as an input.  
In order to make contact with experiment, it will be necessary to shift to a
scheme in which the more precisely measured $G_F$ is instead used as an input.
In this language, the standard-model weak mixing angle $s_Z$ is defined in terms of $\alpha$, $M_Z$ and
$G_F$ as 
\begin{equation}
  c_Z^2 ( 1 - c_Z^2) \equiv \dfrac{e^2}{4\sqrt{2} G_F M_Z^2}, \qquad
  s_Z^2 \equiv 1 - c_Z^2.
\label{eq:Z_scheme}
\end{equation}
The relationship between $\alpha$, $G_F$ and $M_Z$ is altered
by the non-standard elements of our model;  thus 
the weak mixing angle appearing in the amplitude for four-fermion processes
is shifted from its standard model value
\begin{equation}
  c^2 = c_Z^2 + \Delta_Z.
\label{eq:Delta_Z}
\end{equation}
The size of $M_W$ relative to $M_Z$ will likewise be shifted from its standard model value.  We will calculate the values of $\Delta_Z$ and $M_W$ separately for each isospin limit considered below.  

\subsection{Moderate Isospin Violation}

 In this section, we will assume the small, but non-zero
isospin splitting
\begin{equation}
{M^2_{W\hat{n}}\ {\rm or}\ {\cal M}^2_{\hat m}} \gg \Delta M^2_{Z\hat{k}} \gg
(M^2_Z - M^2_W){M^2_W \over {M^2_{W\hat{n}}\ {\rm or}\ {\cal M}^2_{\hat m}}}~,
\end{equation}
for all heavy vector mesons. An examination of eqns.  (\ref{eq:Z_ell_residue_WY}), 
(\ref{eq:Z_ell_residue_YY}),  and (\ref{eq:Z_ell_residue_WW}) for the $\xi_{Z\hat{\ell}}$  shows that the exchange of heavy 
resonances is negligible in this limit.  The only corrections to electroweak 
processes at ``low energies'', {\it i.e.} energies much less than the heavy boson masses,
arise from corrections to the $W$ and $Z$ couplings and masses; all
electroweak corrections are oblique.

Since all corrections are oblique, we may summarize the amplitude for
low-energy four-fermion neutral weak current processes by
\begin{equation}
  {\cal M}_{\rm NC}
  = e^2 \dfrac{{\cal Q} {\cal Q}'}{Q^2}
     +\dfrac{(I_3-s^2 {\cal Q}) (I'_3 - s^2 {\cal Q}')}
            {\left(\dfrac{s^2 c^2}{e^2}-\dfrac{S}{16\pi}\right)Q^2
             +\dfrac{1}{4\sqrt{2} G_F} - s^2 c^2 M_Z^2 \dfrac{T}{4\pi}
            },
\label{eq:NC1}
\end{equation}
and charged-current processes by
\begin{eqnarray}
  {\cal M}_{\rm CC}
  =  \dfrac{(I_{+} I'_{-} + I_{-} I'_{+})/2}
             {\left(\dfrac{s^2}{e^2}-\dfrac{S+U}{16\pi}\right)Q^2
             +\dfrac{1}{4\sqrt{2} G_F}
            }~.
\label{eq:CC1}
\end{eqnarray}
Here $I^{(\prime)}_a$ and ${\cal Q}^{(\prime)}$ are weak-isospin and charge
of the corresponding fermion, and the weak mixing angle is denoted by $s^2$ ($c^2\equiv 1-s^2$).
$S$, $T$, and $U$ are the familiar oblique
electroweak parameters. Given the form of eqn. (\ref{eq:NC1}) and the constraint from the
low-energy $\rho$-parameter (eqn. (\ref{eq:rho-parameter})), we expect to find $T = 0$.

Eqs.(\ref{eq:NC1}) and (\ref{eq:CC1}) may be compared with
our definitions (neglecting the heavy boson contributions) in eqns. (\ref{eq:NC_WW}), (\ref{eq:NC_WY}), (\ref{eq:NC_YY}),
and (\ref{eq:CC_WW}) to obtain the pole residues
\begin{eqnarray}
  {}[\xi_Z]_{WW}
  &=& \dfrac{c^4}{\dfrac{s^2c^2}{e^2}-\dfrac{S}{16\pi}}, 
 \nonumber  \\
  {}[\xi_Z]_{WY}
  &=& \dfrac{-s^2 c^2}{\dfrac{s^2c^2}{e^2}-\dfrac{S}{16\pi}}, 
 \nonumber \\
  {}[\xi_Z]_{YY}
  &=& \dfrac{s^4}{\dfrac{s^2c^2}{e^2}-\dfrac{S}{16\pi}}, 
 \label{eq:xis}  \\
  {}[\xi_W]_{WW}
  &=& \dfrac{1}{\dfrac{s^2}{e^2}-\dfrac{S+U}{16\pi}}, 
\nonumber
\end{eqnarray}
and the masses of the weak gauge bosons 
\begin{eqnarray}
  M_Z^2 &=& \dfrac{1}{4\sqrt{2} G_F 
    \left(\dfrac{s^2 c^2}{e^2} - \dfrac{S}{16\pi} 
    + s^2 c^2 \dfrac{T}{4\pi}
    \right)}, 
\label{eq:M_Z1}
  \\
  M_W^2 &=& \dfrac{1}{4\sqrt{2} G_F 
    \left(\dfrac{s^2}{e^2} - \dfrac{S+U}{16\pi}
    \right)}, 
\label{eq:M_W1}
\end{eqnarray}
where we have used ${\cal Q}^{(\prime)}=I^{(\prime)}_3 + Y^{(\prime)}$.
The pole residues ($[\xi_Z]_{WW}$, $[\xi_Z]_{WY}$, $[\xi_Z]_{YY}$) can 
be precisely determined from the LEP/SLC precision measurements on the
Z-pole, since they can be related to the decay widths and the
asymmetries of the $Z$ boson.

Comparing equations (\ref{eq:Z_scheme}),  (\ref{eq:Delta_Z}) and (\ref{eq:M_Z1}) gives us the value of $\Delta_Z$ in this isospin limit,
\begin{equation}
  \Delta_Z = \dfrac{\alpha}{c_Z^2 - s_Z^2}\left[
    - \frac{1}{4} S + s_Z^2 c_Z^2 T
  \right].
\end{equation}
Knowing the shift in the weak mixing angle allows us to find the $W$ mass and to rewrite the pole residues of the $Z$ boson in terms of the tree-level standard model weak angle ($c_Z$).
The $W$ boson mass is calculated from Eq.(\ref{eq:M_W1}),
\begin{equation}
  M_W^2 = c_Z^2 M_Z^2 \left[
    1 + \dfrac{\alpha}{c_Z^2 - s_Z^2}\left[
         -\frac{1}{2} S + c_Z^2 T + \dfrac{c_Z^2 - s_Z^2}{4s_Z^2} U
        \right]
  \right].
\label{eq:Z_scheme_MW}
\end{equation}
The pole residues of the $Z$ boson are given by
\begin{eqnarray}
  \dfrac{1}{e^2}[\xi_Z]_{WW}
  &=& \dfrac{c_Z^2}{s_Z^2} 
     +\dfrac{\alpha}{s_Z^2(c_Z^2 - s_Z^2)}
      \left[
        - \frac{1}{2} S + c_Z^2 T
      \right],
\label{eq:Z_scheme_WW}
  \\
  \dfrac{1}{e^2}[\xi_Z]_{WY}
  &=& 
    -1 -  \dfrac{\alpha}{4s_Z^2 c_Z^2} S.
\label{eq:Z_scheme_WY}
\end{eqnarray}

We can obtain expressions for $S$, $T$, and $U$ by comparing the
residues as calculated here with their values as calculated earlier in terms of the gauge boson spectrum.  To do this self-consistently, we must first insert our calculated value of $M_W$ (\ref{eq:Z_scheme_MW}) into the earlier expressions  (\ref{eq:Z_residue_WW}) and (\ref{eq:Z_residue_WY}) for  $[\xi_Z]_{WW}$ and $[\xi_Z]_{WY}$; this will align them with our current scheme of using $G_F$ rather than $M_W$ as an experimental input.  The result of these calculations is 
\begin{eqnarray}
  \alpha S 
  &=& 4 s_Z^2 c_Z^2 M_Z^2 
      \left( \Sigma_Z - \Sigma_{\cal M} \right)~,
 \nonumber  \\
  \alpha T
  &=& s_Z^2 M_Z^2 
 \left( \Sigma_Z - \Sigma_W - \Sigma_{\cal M}\right)~,
\label{eq:STU}
 \\
  \alpha U
  &=& 0,
\nonumber
\end{eqnarray}
to lowest nontrivial order in $1/M^2_{Z\hat{k}}$, where we have
defined the sums
\begin{equation}
\Sigma_{Z} \equiv \sum_{\hat{k}=1}^K {1\over M^2_{Z\hat{k}}}\ ,\ \ \ \ \ 
\Sigma_{W} \equiv  \sum_{\hat{n}=1}^N {1\over M^2_{W\hat{n}}}\ ,\ \ \ \ \ 
\Sigma_{\cal M}  \equiv \sum_{\hat{m}=1}^M {1\over {\cal M}^2_{\hat{m}}}.
\end{equation}
However, for the assumed level of isospin violation $\Delta M^2_{Z\hat{k}}
\ll \{ M^2_{W{\hat{n}}}\ {\rm or}\ M^2_{\hat{m}} \}$, we have
\begin{equation}
\Sigma_Z - \Sigma_W - \Sigma_{\cal M} \approx 0~,
\label{eq:isospin_consequence}
\end{equation}
and hence
\begin{eqnarray}
  \alpha S 
  & \approx & 4 s_Z^2 c_Z^2 M_Z^2\, \Sigma_W~,
        \\
  \alpha T
  & \approx & 0~.
  \end{eqnarray}

As a check of the consistency of these results, we may use the explicit form of
the couplings of the $Z$-boson (\ref{eq:zzcoup}) to see if $Z$-exchange accounts for the low-energy
neutral current relevant to atomic parity violation, eqn. (\ref{eq:APV}). We find that,
so long as eqn. (\ref{eq:isospin_consequence}) holds, $Z$-exchange suffices to reproduce 
eqn. (\ref{eq:APV}).

\subsection{Large Isospin Violation}

We next consider the case in which the isospin violation in the spectrum of heavy resonance
masses is large
\begin{equation}
\Delta M^2_{Z\hat{k}} \simeq \{ M^2_{W\hat{n}}\ {\rm or}\ {\cal M}^2_{\hat{m}}\}~.
\end{equation}
In this case, examining eqn. (\ref{eq:Z_ell_residue_YY}), we expect
that the exchange of heavy neutral resonances may be important and result in an extra four-fermion
contribution at low-energies proportional to $J^\mu_Y J_{Y\mu}$. Given the constraint that
the low-energy $\rho$-parameter must equal one, we see that eqn. (\ref{eq:NC1}) must be modified to
\begin{eqnarray}
  {\cal M}_{\rm NC}
  = e^2 \dfrac{{\cal Q} {\cal Q}'}{Q^2}
     & +& \dfrac{(I_3-s^2 {\cal Q}) (I'_3 - s^2 {\cal Q}')}
            {\left(\dfrac{s^2 c^2}{e^2}-\dfrac{S}{16\pi}\right)Q^2
             +\dfrac{1}{4\sqrt{2} G_F} - s^2 c^2 M_Z^2 \dfrac{T}{4\pi}
            } 
\nonumber \\
       &-& 8 G^2_F s^2 c^2 M^2_Z \, T\, \left[{({\cal Q}-I_3)({\cal Q}'-I'_3)\over \pi}\right]~.
\label{eq:NC2}
\end{eqnarray}
The calculations leading to eqns. (\ref{eq:STU}) remain unchanged. Here, however,
we cannot simplify further using eqn. (\ref{eq:isospin_consequence}).
The eqns. (\ref{eq:STU}) imply that
\begin{equation}
\alpha S - 4 c^2_Z\,\alpha T = 4 s^2_Z c^2_Z M^2_Z\,  \Sigma_W ~.
\label{eq:STU-consequence}
\end{equation}

The explicit form of the couplings of the $Z$-boson (\ref{eq:zzcoup}) and the additional
$J^\mu_Y J_{Y\mu}$ contribution in eqn. (\ref{eq:NC2})
together account for the low-energy
neutral current relevant to atomic parity violation, eqn. (\ref{eq:APV}).

\subsection{Small Isospin Violation}

Finally, we consider the case of small isospin violation in the heavy resonance masses
\begin{equation}
\Delta M^2_{Z\hat{k}} \simeq (M^2_Z - M^2_W) {M^2_W \over 
{M^2_{W\hat{n}}\ {\rm or}\ {\cal M}^2_{\hat{m}}}}~.
\end{equation}
We note that this is the {\it minimum} size of isospin violation one expects in the
limit $g_{N+1} \to 0$. In this case, examining eqns. (\ref{eq:W_ell_residue_WW}) and
(\ref{eq:Z_ell_residue_WW}), we expect that the exchange of heavy resonances may
be important and result in an extra four-fermion contribution to both low-energy neutral and charged
current exchange proportional to the square of the weak $SU(2)$ currents 
$\vec{J}^\mu \cdot \vec{J}_{\mu}$. The form of amplitudes for neutral-current processes are then
given by
\begin{equation}
  {\cal M}_{\rm NC}
  = e^2 \dfrac{{\cal Q} {\cal Q}'}{Q^2}
     +\dfrac{(I_3-s^2 {\cal Q}) (I'_3 - s^2 {\cal Q}')}
            {\left(\dfrac{s^2 c^2}{e^2}-\dfrac{S}{16\pi}\right)Q^2
             +\dfrac{1+(e^2 \delta/ 16\pi s^2 c^2)}{4\sqrt{2} G_F}
              }
      + \sqrt{2} G_F \,{e^2 \delta\over 4 \pi s^2 c^2}\, I_3 I'_3~,
\label{eq:NC3}
\end{equation}
and charged-current processes by
\begin{eqnarray}
  {\cal M}_{\rm CC}
  =  \dfrac{(I_{+} I'_{-} + I_{-} I'_{+})/2}
             {\left(\dfrac{s^2}{e^2}-\dfrac{S}{16\pi}\right)Q^2
             +\dfrac{1+(e^2  \delta/16\pi  s^2 c^2)}{4\sqrt{2} G_F}
            }
        + \sqrt{2} G_F\, {e^2  \delta\over 4\pi s^2 c^2} \, {(I_{+} I'_{-} + I_{-} I'_{+}) \over 2}~,
\label{eq:CC3}
\end{eqnarray}
where isospin symmetry implies $T=U=0$, and the dimensionless parameter
$\delta$ measures the size of the contributions from heavy resonance exchange.
Note that, formally, eqns. (\ref{eq:W_ell_residue_WW}) and (\ref{eq:Z_ell_residue_WW})
provide an expression for $\delta$ in terms of the boson masses and $e^2$. 

The expressions for the pole residues in eqns. (\ref{eq:xis}) remain the same, while
for the gauge-boson masses we find
\begin{eqnarray}
  M_Z^2 &=& \dfrac{1}{4\sqrt{2} G_F 
    \left(\dfrac{s^2 c^2}{e^2} - \dfrac{S+\delta}{16\pi} 
    \right)}, 
\label{eq:M_Z3}
  \\
  M_W^2 &=& \dfrac{1}{4\sqrt{2} G_F 
    \left(\dfrac{s^2}{e^2} - \dfrac{S+\delta/c^2}{16\pi}
    \right)}~.
\label{eq:M_W3}
\end{eqnarray}
From eqns. (\ref{eq:Z_scheme}) and (\ref{eq:Delta_Z}), we obtain
\begin{equation}
\Delta_Z = -\, {\alpha \over 4(c^2_Z-s^2_Z)}\, (S+\delta)~,
\end{equation}
and consequently, from eqn. (\ref{eq:M_W3})
\begin{equation}
M^2_W = c^2_Z  M^2_Z \left[
1+{\alpha \over c^2_Z-s^2_Z} 
\left[
-{1\over 2} S
-{\delta \over 4 c^2_Z}
\right]
\right]~.
\label{eq:M_W4}
\end{equation}

To this order, we continue to find, from analyzing $[\xi_Z]_{WY}$ for example,
\begin{equation}
 \alpha S = 4 s_Z^2 c_Z^2 M_Z^2  \left( \Sigma_Z - \Sigma_{\cal M} \right) =
 4 s^2_Z c^2_Z M^2_Z \Sigma_W~,
 \end{equation}
where the last equality follows from the isospin invariance of the heavy spectrum.
In terms of $M^2_W$, $\delta$ is determined by eqn. (\ref{eq:M_W4})
\begin{equation}
\alpha \delta = 4 c^2_Z (c^2_Z-s^2_Z) \left(1-{M^2_W \over c^2_Z M^2_Z}\right)
- 8 s^2_Z c^4_Z M^2_Z (\Sigma_Z - \Sigma_{\cal M})~.
\end{equation}
In practice, however, $M^2_W$ is not precisely measured and eqn. (\ref{eq:M_W4})
is used as just one constraint in a combined $S$ \& $\delta$ fit.

\section{Discussion and Conclusions}

In this paper, using
deconstruction, we have calculated the form of the corrections to the electroweak interactions in a large class of Higgsless models, allowing for arbitrary 5-D geometry, position-dependent 
gauge coupling, and brane kinetic energy terms. 
We have shown that many models considered in the literature, including those most likely to be
phenomenologically viable, are in this class. By analyzing the asymptotic behavior of
the correlation function of gauge currents at high momentum, we have extracted 
the exact form of the relevant correlation functions at tree-level and computed the
corrections to precision electroweak observables in terms of the spectrum
of heavy vector bosons. Although we have stressed our results as they apply to continuum
5-D models, they apply also to models of extended  electroweak gauge symmetries 
motivated by models of hidden local symmetry.  In a subsequent paper, we will address  the consequences of our results for model-building in light of current  precision electroweak data.

 We have shown that the amount of isospin violation in the
heavy boson spectrum (as measured by $\Delta M^2_{Z\hat{k}}$) determines
when nonoblique corrections due to the interactions of fermions with
the heavy vector bosons become important.   For moderate
$\Delta M^2_{Z\hat{k}}$, exchange of heavy resonances is
negligible.    For large $\Delta M^2_{Z\hat{k}}$, heavy neutral resonance exchange
becomes significant in the $J^\mu_Y J_{Y\mu}$ channel;  for example, light Z-boson
exchange no longer suffices to account for atomic parity violation.
In the case of  small $\Delta M^2_{Z\hat{k}}$, exchange of both charged
and neutral heavy resonances is important; while the small size of isospin
violation makes $T$ and $U$ zero in this limit, and electroweak fits must include
an additional parameter $\delta$, which measures the size of contributions
from heavy resonance exchange.

Independent of the amount of isospin violation, we have found the following constraint\footnote{Note that to the order we are working, it does not matter whether we write $\cos^2\theta_W$ or $c_Z^2$ in this relationship because $S$, $T$ and the sum over $M^{-2}_{W\hat{n}}$ are all small.} on the oblique electroweak parameters $S$ and $T$ 
\begin{equation}
\alpha S - 4 \cos^2\theta_W\,\alpha T = 4 \sin^2\theta_W \cos^2\theta_W M^2_Z\,   \sum_{\hat{n}=1}^N {1\over M^2_{W\hat{n}}} ~,
\label{eq:summary}
\end{equation}
in the class of models\footnote{Recently a Higgsless model with vanishing
$S$ has been proposed \protect{\cite{Casalbuoni:2004id}} -- the simplest version
of which was previously analyzed in \protect{\cite{Chivukula:2003wj}}.  However, the chiral symmetries of this model, like those of QCD-like technicolor, allow for an
${\cal O}(p^4)$ term (not considered in  \protect{\cite{Casalbuoni:2004id}})
which directly contributes to the $S$ parameter.  When this term is included, the value of $S$ is expected to
be of order 1.} studied here. In the limit of moderate or small isospin violation, one finds $T\approx 0$, which simplifies this relationship.   In any unitary theory \cite{SekharChivukula:2001hz, Chivukula:2002ej}, we expect the mass of the lightest
vector $m_{W\hat{1}}$ to be less than $\sqrt{8 \pi} v$ ($v\approx 246$ GeV) -- the scale at
which $WW$ spin-0 isospin-0 elastic scattering would violate unitarity in the 
standard model in the absence of a higgs
boson \cite{Dicus:1973vj,Cornwall:1973tb, Cornwall:1974km,Lee:1977yc, Lee:1977eg,Veltman:1977rt}.
Evaluating eqn. (\ref{eq:summary}) we see that we expect 
$S-4 \cos^2\theta_W T$ to be of order one-half  or larger, generalizing the result of \cite{Chivukula:2004kg}.

It is interesting to note how this bound applies to the model discussed in \cite{Cacciapaglia:2004jz},
the deconstructed version of which is sketched in Fig. 4. 
In the case with a $U(1)_{B-L}$ TeV brane kinetic term, these authors find
\begin{eqnarray}
S&\approx&\frac{ 6 \pi }{g^2\log \frac{R'}{R}}-\frac{ 8 \pi }{g^2} 
\left( 1-\left(\frac{g'}{g}\right)^2 \right) \frac{\tau'^2}{(R \log 
R'/R)^2}~, \label{eq:Staup}\\
T&\approx&-\frac{ 2 \pi }{g^2} \left( 1-\left(\frac{g'}{g}\right)^4 
\right) \frac{\tau'^2}{(R \log R'/R)^2}~,\\
U&\approx&0~,
\end{eqnarray}
where $g$ and $g'$ are the usual $SU(2)_W$ and $U(1)_Y$ couplings, $\tau'$ is
a measure of the strength of the brane kinetic term, and $R$ and $R'$ are the positions
of the Planck and TeV branes respectively.
Noting that $1+(g'/g)^2 \approx 1/{\cos^2\theta_W}$, we immediately find
\begin{equation}
S-4 \cos^2\theta_W T \approx {6\pi \over g^2 \log{R'\over R}} ~.
\label{eq:finalrel}
\end{equation}
To ensure unitarity, one expects $1/R' \simeq 1\, {\rm TeV}$; using $1/R \simeq M_{planck}$ one obtains $S-4 \cos^2\theta_W  T
= {\cal O}(1)$ as expected.  Note that although the individual values of $S$ and $T$ depend on $\tau'$, the relationship between them in eqn. (\ref{eq:finalrel})   agrees with our result (\ref{eq:summary}) {\it independent} of the strength of the brane kinetic energy term.

 \acknowledgments

We would like to thank Nick Evans, Takeo Moroi, Yasuhiro Shimizu, and
John Terning for discussions.
M.K. acknowledges support by the 21st Century COE Program of Nagoya University 
provided by JSPS (15COEG01). M.T.'s work is supported in part by the JSPS Grant-in-Aid for Scientific Research No.16540226. H.J.H. is supported by the US Department of Energy grant
DE-FG03-93ER40757.

\appendix



\begin{thebibliography}{99}

\bibitem{Csaki:2003dt}
C.~Csaki, C.~Grojean, H.~Murayama, L.~Pilo, and J.~Terning, {\it Gauge theories
  on an interval: Unitarity without a higgs},
  \href{http://xxx.lanl.gov/abs/hep-ph/0305237}{{\tt hep-ph/0305237}}.

\bibitem{SekharChivukula:2001hz}
R.~Sekhar~Chivukula, D.~A. Dicus, and H.-J. He, {\it Unitarity of compactified
  five dimensional yang-mills theory},  {\em Phys. Lett.} {\bf B525} (2002)
  175--182, [\href{http://xxx.lanl.gov/abs/hep-ph/0111016}{{\tt
  hep-ph/0111016}}].

\bibitem{Chivukula:2002ej}
R.~S. Chivukula and H.-J. He, {\it Unitarity of deconstructed five-dimensional
  yang-mills theory},  {\em Phys. Lett.} {\bf B532} (2002) 121--128,
  [\href{http://xxx.lanl.gov/abs/hep-ph/0201164}{{\tt hep-ph/0201164}}].

\bibitem{Chivukula:2003kq}
R.~S. Chivukula, D.~A. Dicus, H.-J. He, and S.~Nandi, {\it Unitarity of the
  higher dimensional standard model},  {\em Phys. Lett.} {\bf B562} (2003)
  109--117, [\href{http://xxx.lanl.gov/abs/hep-ph/0302263}{{\tt
  hep-ph/0302263}}].

\bibitem{Higgs:1964ia}
P.~W. Higgs, {\it Broken symmetries, massless particles and gauge fields},
  {\em Phys. Lett.} {\bf 12} (1964) 132--133.

\bibitem{Maldacena:1998re}
J.~M. Maldacena, {\it The large n limit of superconformal field theories and
  supergravity},  {\em Adv. Theor. Math. Phys.} {\bf 2} (1998) 231--252,
  [\href{http://xxx.lanl.gov/abs/hep-th/9711200}{{\tt hep-th/9711200}}].

\bibitem{Gubser:1998bc}
S.~S. Gubser, I.~R. Klebanov, and A.~M. Polyakov, {\it Gauge theory correlators
  from non-critical string theory},  {\em Phys. Lett.} {\bf B428} (1998)
  105--114, [\href{http://xxx.lanl.gov/abs/hep-th/9802109}{{\tt
  hep-th/9802109}}].

\bibitem{Witten:1998qj}
E.~Witten, {\it Anti-de sitter space and holography},  {\em Adv. Theor. Math.
  Phys.} {\bf 2} (1998) 253--291,
  [\href{http://xxx.lanl.gov/abs/hep-th/9802150}{{\tt hep-th/9802150}}].

\bibitem{Aharony:1999ti}
O.~Aharony, S.~S. Gubser, J.~M. Maldacena, H.~Ooguri, and Y.~Oz, {\it Large n
  field theories, string theory and gravity},  {\em Phys. Rept.} {\bf 323}
  (2000) 183--386, [\href{http://xxx.lanl.gov/abs/hep-th/9905111}{{\tt
  hep-th/9905111}}].

\bibitem{Weinberg:1979bn}
S.~Weinberg, {\it Implications of dynamical symmetry breaking: An addendum},
  {\em Phys. Rev.} {\bf D19} (1979) 1277--1280.

\bibitem{Susskind:1979ms}
L.~Susskind, {\it Dynamics of spontaneous symmetry breaking in the weinberg-
  salam theory},  {\em Phys. Rev.} {\bf D20} (1979) 2619--2625.

\bibitem{Holdom:1981rm}
B.~Holdom, {\it Raising the sideways scale},  {\em Phys. Rev.} {\bf D24} (1981)
  1441.

\bibitem{Holdom:1985sk}
B.~Holdom, {\it Techniodor},  {\em Phys. Lett.} {\bf B150} (1985) 301.

\bibitem{Yamawaki:1986zg}
K.~Yamawaki, M.~Bando, and K.-i. Matumoto, {\it Scale invariant technicolor
  model and a technidilaton},  {\em Phys. Rev. Lett.} {\bf 56} (1986) 1335.

\bibitem{Appelquist:1986an}
T.~W. Appelquist, D.~Karabali, and L.~C.~R. Wijewardhana, {\it Chiral
  hierarchies and the flavor changing neutral current problem in technicolor},
  {\em Phys. Rev. Lett.} {\bf 57} (1986) 957.

\bibitem{Appelquist:1987tr}
T.~Appelquist and L.~C.~R. Wijewardhana, {\it Chiral hierarchies and chiral
  perturbations in technicolor},  {\em Phys. Rev.} {\bf D35} (1987) 774.

\bibitem{Appelquist:1987fc}
T.~Appelquist and L.~C.~R. Wijewardhana, {\it Chiral hierarchies from slowly
  running couplings in technicolor theories},  {\em Phys. Rev.} {\bf D36}
  (1987) 568.

\bibitem{Arkani-Hamed:2001ca}
N.~Arkani-Hamed, A.~G. Cohen, and H.~Georgi, {\it (de)constructing dimensions},
   {\em Phys. Rev. Lett.} {\bf 86} (2001) 4757--4761,
  [\href{http://xxx.lanl.gov/abs/hep-th/0104005}{{\tt hep-th/0104005}}].

\bibitem{Hill:2000mu}
C.~T. Hill, S.~Pokorski, and J.~Wang, {\it Gauge invariant effective lagrangian
  for kaluza-klein modes},  {\em Phys. Rev.} {\bf D64} (2001) 105005,
  [\href{http://xxx.lanl.gov/abs/hep-th/0104035}{{\tt hep-th/0104035}}].

\bibitem{Perelstein:2004sc}
M.~Perelstein,
arXiv:hep-ph/0408072.

\bibitem{Cacciapaglia:2004jz}
G.~Cacciapaglia, C.~Csaki, C.~Grojean, and J.~Terning, {\it Oblique corrections
  from higgsless models in warped space},
  \href{http://xxx.lanl.gov/abs/hep-ph/0401160}{{\tt hep-ph/0401160}}.

\bibitem{Son:2003et}
D.~T. Son and M.~A. Stephanov, {\it Qcd and dimensional deconstruction},
  \href{http://xxx.lanl.gov/abs/hep-ph/0304182}{{\tt hep-ph/0304182}}.

\bibitem{Hirn:2004ze}
J.~Hirn and J.~Stern, {\it The role of spurions in higgs-less electroweak
  effective theories},  \href{http://xxx.lanl.gov/abs/hep-ph/0401032}{{\tt
  hep-ph/0401032}}.

\bibitem{Chivukula:2004kg}
R.~S. Chivukula, M.~Kurachi, and M.~Tanabashi, {\it Generalized weinberg sum
  rules in deconstructed qcd},
  \href{http://xxx.lanl.gov/abs/hep-ph/0403112}{{\tt hep-ph/0403112}}.

\bibitem{Peskin:1992sw}
M.~E. Peskin and T.~Takeuchi, {\it Estimation of oblique electroweak
  corrections},  {\em Phys. Rev.} {\bf D46} (1992) 381--409.

\bibitem{Csaki:2003zu}
C.~Csaki, C.~Grojean, L.~Pilo, and J.~Terning, {\it Towards a realistic model
  of higgsless electroweak symmetry breaking},  {\em Phys. Rev. Lett.} {\bf 92}
  (2004) 101802, [\href{http://xxx.lanl.gov/abs/hep-ph/0308038}{{\tt
  hep-ph/0308038}}].

\bibitem{Nomura:2003du}
Y.~Nomura, {\it Higgsless theory of electroweak symmetry breaking from warped
  space},  {\em JHEP} {\bf 11} (2003) 050,
  [\href{http://xxx.lanl.gov/abs/hep-ph/0309189}{{\tt hep-ph/0309189}}].

\bibitem{Barbieri:2003pr}
R.~Barbieri, A.~Pomarol, and R.~Rattazzi, {\it Weakly coupled higgsless
  theories and precision electroweak tests},
  \href{http://xxx.lanl.gov/abs/hep-ph/0310285}{{\tt hep-ph/0310285}}.

\bibitem{Davoudiasl:2003me}
H.~Davoudiasl, J.~L. Hewett, B.~Lillie, and T.~G. Rizzo, {\it Higgsless
  electroweak symmetry breaking in warped backgrounds: Constraints and
  signatures},  \href{http://xxx.lanl.gov/abs/hep-ph/0312193}{{\tt
  hep-ph/0312193}}.

\bibitem{Burdman:2003ya}
G.~Burdman and Y.~Nomura, {\it Holographic theories of electroweak symmetry
  breaking without a higgs boson},
  \href{http://xxx.lanl.gov/abs/hep-ph/0312247}{{\tt hep-ph/0312247}}.

\bibitem{Davoudiasl:2004pw}
H.~Davoudiasl, J.~L. Hewett, B.~Lillie, and T.~G. Rizzo, {\it Warped higgsless
  models with ir-brane kinetic terms},  {\em JHEP} {\bf 05} (2004) 015,
  [\href{http://xxx.lanl.gov/abs/hep-ph/0403300}{{\tt hep-ph/0403300}}].

\bibitem{Barbieri:2004qk}
R.~Barbieri, A.~Pomarol, R.~Rattazzi, and A.~Strumia, {\it Electroweak symmetry
  breaking after lep1 and lep2},
  \href{http://xxx.lanl.gov/abs/hep-ph/0405040}{{\tt hep-ph/0405040}}.

\bibitem{Foadi:2003xa}
R.~Foadi, S.~Gopalakrishna, and C.~Schmidt, {\it Higgsless electroweak symmetry
  breaking from theory space},  {\em JHEP} {\bf 03} (2004) 042,
  [\href{http://xxx.lanl.gov/abs/hep-ph/0312324}{{\tt hep-ph/0312324}}].

\bibitem{Chivukula:2003wj}
R.~S. Chivukula, H.-J. He, J.~Howard, and E.~H. Simmons, {\it The structure of
  electroweak corrections due to extended gauge symmetries},  {\em Phys. Rev.}
  {\bf D69} (2004) 015009, [\href{http://xxx.lanl.gov/abs/hep-ph/0307209}{{\tt
  hep-ph/0307209}}].

\bibitem{Casalbuoni:1985kq}
R.~Casalbuoni, S.~De~Curtis, D.~Dominici, and R.~Gatto, {\it Effective weak
  interaction theory with possible new vector resonance from a strong higgs
  sector},  {\em Phys. Lett.} {\bf B155} (1985) 95.

\bibitem{Casalbuoni:1996qt}
R.~Casalbuoni {\em et.~al.}, {\it Degenerate bess model: The possibility of a
  low energy strong electroweak sector},  {\em Phys. Rev.} {\bf D53} (1996)
  5201--5221, [\href{http://xxx.lanl.gov/abs/hep-ph/9510431}{{\tt
  hep-ph/9510431}}].

\bibitem{Bando:1985ej}
M.~Bando, T.~Kugo, S.~Uehara, K.~Yamawaki, and T.~Yanagida, {\it Is rho meson a
  dynamical gauge boson of hidden local symmetry?},  {\em Phys. Rev. Lett.}
  {\bf 54} (1985) 1215.

\bibitem{Bando:1985rf}
M.~Bando, T.~Kugo, and K.~Yamawaki, {\it On the vector mesons as dynamical
  gauge bosons of hidden local symmetries},  {\em Nucl. Phys.} {\bf B259}
  (1985) 493.

\bibitem{Bando:1988ym}
M.~Bando, T.~Fujiwara, and K.~Yamawaki, {\it Generalized hidden local symmetry
  and the a1 meson},  {\em Prog. Theor. Phys.} {\bf 79} (1988) 1140.

\bibitem{Bando:1988br}
M.~Bando, T.~Kugo, and K.~Yamawaki, {\it Nonlinear realization and hidden local
  symmetries},  {\em Phys. Rept.} {\bf 164} (1988) 217--314.

\bibitem{Harada:2003jx}
M.~Harada and K.~Yamawaki, {\it Hidden local symmetry at loop: A new
  perspective of composite gauge boson and chiral phase transition},  {\em
  Phys. Rept.} {\bf 381} (2003) 1--233,
  [\href{http://xxx.lanl.gov/abs/hep-ph/0302103}{{\tt hep-ph/0302103}}].

\bibitem{Georgi:1986hf}
H.~Georgi, {\it A tool kit for builders of composite models},  {\em Nucl.
  Phys.} {\bf B266} (1986) 274.

\bibitem{Hebecker:2001jb}
A.~Hebecker and J.~March-Russell, {\it The structure of gut breaking by
  orbifolding},  {\em Nucl. Phys.} {\bf B625} (2002) 128--150,
  [\href{http://xxx.lanl.gov/abs/hep-ph/0107039}{{\tt hep-ph/0107039}}].

\bibitem{Agashe:2003zs}
K.~Agashe, A.~Delgado, M.~J. May, and R.~Sundrum, {\it Rs1, custodial isospin
  and precision tests},  {\em JHEP} {\bf 08} (2003) 050,
  [\href{http://xxx.lanl.gov/abs/hep-ph/0308036}{{\tt hep-ph/0308036}}].

\bibitem{Casalbuoni:2004id}
R.~Casalbuoni, S.~De~Curtis, and D.~Dominici, {\it Moose models with vanishing
  s parameter},  \href{http://xxx.lanl.gov/abs/hep-ph/0405188}{{\tt
  hep-ph/0405188}}.

\bibitem{Dicus:1973vj}
D.~A. Dicus and V.~S. Mathur, {\it Upper bounds on the values of masses in
  unified gauge theories},  {\em Phys. Rev.} {\bf D7} (1973) 3111--3114.

\bibitem{Cornwall:1973tb}
J.~M. Cornwall, D.~N. Levin, and G.~Tiktopoulos, {\it Uniqueness of
  spontaneously broken gauge theories},  {\em Phys. Rev. Lett.} {\bf 30} (1973)
  1268--1270.

\bibitem{Cornwall:1974km}
J.~M. Cornwall, D.~N. Levin, and G.~Tiktopoulos, {\it Derivation of gauge
  invariance from high-energy unitarity bounds on the s - matrix},  {\em Phys.
  Rev.} {\bf D10} (1974) 1145.

\bibitem{Lee:1977yc}
B.~W. Lee, C.~Quigg, and H.~B. Thacker, {\it The strength of weak interactions
  at very high-energies and the higgs boson mass},  {\em Phys. Rev. Lett.} {\bf
  38} (1977) 883.

\bibitem{Lee:1977eg}
B.~W. Lee, C.~Quigg, and H.~B. Thacker, {\it Weak interactions at very
  high-energies: The role of the higgs boson mass},  {\em Phys. Rev.} {\bf D16}
  (1977) 1519.

\bibitem{Veltman:1977rt}
M.~J.~G. Veltman, {\it Second threshold in weak interactions},  {\em Acta Phys.
  Polon.} {\bf B8} (1977) 475.



\end{thebibliography}

\end{document}